\documentclass[10pt,journal,compsoc]{IEEEtran}
\usepackage{tabularx}
\usepackage{booktabs}
\usepackage{multirow}
\usepackage{amssymb}
\usepackage{bm}
\usepackage[ruled,vlined,linesnumbered]{algorithm2e}
\usepackage{placeins}

\newtheorem{example}{Example}
\newtheorem{definition}{Definition}

\ifCLASSOPTIONcompsoc
  \usepackage[nocompress]{cite}
\else
  \usepackage{cite}
\fi
\ifCLASSINFOpdf
   \usepackage[pdftex]{graphicx}
\else
\fi
\usepackage{amsmath}
\usepackage{url}

\begin{document}
%
\title{LEMON: Explainable Entity Matching}
%
%
%
%

\author{Nils~Barlaug
\IEEEcompsocitemizethanks{\IEEEcompsocthanksitem Nils Barlaug is with the Department
of Computer Science, Norwegian University of Science and Technology, Trondheim, Norway and Cognite.\protect\\
E-mail: nils.barlaug@ntnu.no}
}

%
%

\markboth{IEEE Transactions on Knowledge and Data Engineering, 2022}%
{LEMON: Explainable Entity Matching}
%



\IEEEtitleabstractindextext{%
\begin{abstract}
State-of-the-art entity matching (EM) methods are hard to interpret,
and there is significant value in bringing explainable AI to EM.
Unfortunately, most popular explainability methods do not work well out of the box for EM and need adaptation.
In this paper,
we identify three challenges of applying local post hoc feature attribution methods to entity matching:
cross-record interaction effects,
non-match explanations,
and variation in sensitivity.
We propose our novel model-agnostic and schema-flexible method LEMON that addresses all three challenges by
(i) producing dual explanations to avoid cross-record interaction effects,
(ii) introducing the novel concept of attribution potential to explain how two records could have matched,
and (iii) automatically choosing explanation granularity to match the sensitivity of the matcher and record pair in question.
Experiments on public datasets demonstrate that the proposed method is more faithful to the matcher and does a better job of helping users understand the decision boundary of the matcher than previous work.
Furthermore, user studies show that the rate at which human subjects can construct counterfactual examples after seeing an explanation from our proposed method increases from 54\% to 64\% for matches and from 15\% to 49\% for non-matches compared to explanations from a standard adaptation of LIME.
\end{abstract}

}

\maketitle

\IEEEdisplaynontitleabstractindextext

%
\IEEEpeerreviewmaketitle

\IEEEraisesectionheading{\section{Introduction}\label{sec:introduction}}

\IEEEPARstart{E}{ntity} matching is an essential task in data integration \cite{doanPrinciplesDataIntegration2012}.
It is the task of identifying which records refer to the same real-world entity.
Figure~\ref{fig:record-pair} shows an example.
Machine learning has become a standard tool to tackle the variety of data and to avoid laborsome feature engineering from experts while still achieving high accuracy (e.g., \cite{kondaMagellanBuildingEntity2016, mudgalDeepLearningEntity2018, liDeepEntityMatching2020}).
Unfortunately, this is often at the cost of reduced transparency and interpretability.
While it is possible to carefully select classical machine learning methods that are intrinsically interpretable and combine them with classical string similarity metrics,
current state-of-the-art consists of large deep learning models \cite{brunnerEntityMatchingTransformer2020,ebraheemDistributedRepresentationsTuples2018,liDeepEntityMatching2020,mudgalDeepLearningEntity2018}, which offer limited interpretability out of the box.
The possible benefits of being able to explain black-box models are numerous.
To mention some:
1) Researchers can gain new insight into their models and find ways to improve them
2) Practitioners will have a valuable tool for verifying that the models work as expected and debugging those which do not 
3) Companies can gain the necessary transparency they need to trust such black-boxes for mission-critical data integration efforts
4) End-users can be reassured that models and their results can be trusted, or discover themselves that they should not be.

The challenge of explaining machine learning models and their potential benefits is not unique to entity matching.
Therefore, explainable machine learning has in recent years received significant attention from the broader research community \cite{adadiPeekingBlackBoxSurvey2018,gilpinExplainingExplanationsOverview2018,guidottiSurveyMethodsExplaining2019}.
The result is a multitude of techniques and methods with different strengths and weaknesses.
But as previous work has discussed,
applying these techniques to entity matching is non-trivial.
It is necessary to adapt and evolve them to tackle the unique characteristics of entity matching \cite{diciccoInterpretingDeepLearning2019,ebaidEXPLAINEREntityResolution2019,thirumuruganathanExplainingEntityResolution2019}.

\begin{figure}
    \centering
    \scriptsize
    \setlength{\tabcolsep}{4pt}
    \begin{tabularx}{0.48\columnwidth}{l>{\raggedright\arraybackslash}X}
    \toprule
        \textbf{\texttt{title}}    & \texttt{belkin shield micra for ipod touch tint} \\ \addlinespace[0.5em]
        \textbf{\texttt{category}} & \texttt{mp3 accessories} \\ \addlinespace[0.5em]
        \textbf{\texttt{brand}}    & \texttt{belkin} \\ \addlinespace[0.5em]
        \textbf{\texttt{modelno}}  & \texttt{f8z646ttc01} \\ \addlinespace[0.5em]
        \textbf{\texttt{price}}    & \texttt{47.88} \\
    \bottomrule
    \end{tabularx}
    \quad
    \begin{tabularx}{0.48\columnwidth}{l>{\raggedright\arraybackslash}X}
    \toprule
        \textbf{\texttt{title}}    & \texttt{belkin ipod touch shield micra tint-royal purple} \\ \addlinespace[0.5em]
        \textbf{\texttt{category}} & \texttt{cases} \\ \addlinespace[0.5em]
        \textbf{\texttt{brand}}    & \texttt{belkin} \\ \addlinespace[0.5em]
        \textbf{\texttt{modelno}}  & \texttt{f8z646ttc02} \\ \addlinespace[0.5em]
        \textbf{\texttt{price}}    & \texttt{12.49} \\
    \bottomrule
    \end{tabularx}
    \caption{
        Example of two records that need to be classified as either a match or a non-match from the Walmart-Amazon dataset.
        In this case, the records refer to almost the same product --- the only definitive difference being the color.
    }
    \label{fig:record-pair}
\end{figure}

Local post hoc feature attribution methods are perhaps the most popular type of explainability method in general,
and the most studied so far for entity matching~\cite{ebaidEXPLAINEREntityResolution2019,thirumuruganathanExplainingEntityResolution2019,diciccoInterpretingDeepLearning2019,baraldiUsingLandmarksExplaining2021}.
Previous work on explainable entity matching base their work on LIME~\cite{ribeiroWhyShouldTrust2016} --- one of the most popular methods of that type.
In this paper,
we choose to focus mainly on LIME to be consistent with, and for ease of comparison to, earlier work.
Our work is relevant beyond LIME,
and we will reference and include other methods in our experiments,
but we consider in-depth adaptation and treatment of other methods outside of the scope of this paper and hope to address them in future work.

\textit{Challenges}.
Unfortunately, standard local post hoc attribution methods do not work satisfactorily for entity matching out of the box.
We identify three challenges of applying them:
\begin{enumerate}
    \item \textbf{Cross-record interaction effects}: \label{challenge-interaction}
    Since EM is a matching problem, features across a record pair will tend to have strong interaction effects,
    but linear surrogate models such as in LIME implicitly assume independent features.
    This can severely impair the accuracy of the surrogate model.
    \item \textbf{Non-match explanations}: \label{challenge-non-matches}
    In essence, most feature attribution methods analyze the effect of removing features to determine their attribution.
    However, for record pairs for which the matcher is fairly confident that they do not match, so the match score is close to zero, it is unlikely that removal of features will make any significant difference on the match score.
    The result is that we have no significant attributions to explain why the records do not match.
    This is especially important since most record pairs do not match.
    \item \textbf{Variation in sensitivity}: \label{challenge-variation}
    While some record pairs may only need to perturb a few features to trigger a significant change in the output of the matcher,
    others may contain a lot of redundant features,
    making it hard to substantially impact the matching score and provide meaningfully sized attributions.
    It can be hard to make the trade-off between token and attribute level feature granularity.
    One risk being either too fine-grained or unnecessarily course-grained,
    and it differs between specific record pairs in the same dataset, across datasets, and across matchers.
\end{enumerate}
As we will outline in Section~\ref{sec:related-work}, earlier work has only partially addressed these challenges.

\textit{Proposed method}.
Our proposed method addresses all three challenges jointly by:
1) Using dual explanations to avoid cross-record interaction effects.
2) Introducing the novel concept of attribution potential, an improvement over the copy/append perturbation from previous work that is schema-flexible and more robust to dirty data and matchers sensitive to the order of tokens.
3) Choosing an interpretable representation granularity that optimizes the trade-off between counterfactual interpretation and the finest granularity possible.
Our proposed method provides one unified frame of interpretation with the same single explanation format for all record pairs, has no dataset-specific hyperparameters that need tinkering, and does not require matched schemas.
Source code is publicly available\footnote{\url{https://github.com/NilsBarlaug/lemon}}.

Evaluation of explainability methods is still an open problem, and there are no standard ways of evaluating explainable entity matching.
Ideally, in broad terms, we would like to measure to what degree an explanation helps users understand how the model makes a matching decision.
Inspired by the motivations behind counterfactual examples \cite{wachterCounterfactualExplanationsOpening2017},
we argue that a useful attribution explanation should help the user understand where the decision boundary is and what kind of difference in input would be necessary to sway the matcher.
To that end, we propose to ask users what they think would be a minimal change to a record pair to make the matcher change its prediction and then check if they are correct.
We show how this can be done for simulated users as well as human subjects.

Finally, our results show that our proposed method is state-of-the-art,
both in terms of faithfulness and helping users understand the matcher's behavior --- though at the cost of higher runtime.
Additionally, our user study shows great potential for real-world improvement in understanding by human subjects.

\textit{Contributions}.
In summary, our main contributions are:
\begin{itemize}
    \item
    We propose a method that addresses three important challenges of applying local post hoc attribution methods to entity matching:
    1) Cross-record interaction effects, 2) non-match explanation, and 3) variation in sensitivity.
    We show through experiments that this is indeed effective.
    \item
    To evaluate entity matching attribution explanations, we propose a novel evaluation method that aims to measure to what degree explanations help users understand the decision boundary of a matcher. We show how to perform experiments on both simulated users and human subjects.
    \item
    Through extensive experiments on public datasets we show that our proposed method is state-of-the-art both in terms of faithfulness and helping users understand the matcher's behavior.
    We verify the real-world applicability of our proposed method by performing an extensive user study. To the best of our knowledge, we are the first to conduct a user study for explainable entity matching.
\end{itemize}

\textit{Outline}
The rest of the paper is organized as follows.
Section 2 briefly covers related work,
Section 3 covers LIME and its adaptation to entity matching,
and Section 4 goes into the details of our proposed method.
We explain our experimental setup in Section 5, and then we walk through and discuss the experiments in Section 6 before we make our concluding remarks in Section 7.

\section{Related work}\label{sec:related-work}
\textit{Machine learning for EM}.
The immense variety in datasets makes machine learning a natural solution for entity matching.
The traditional approach has been to handcraft string similarity metrics to produce similarity feature vectors and then utilize a classical off-the-shelf machine learning model such as SVM or random forest to classify them \cite{christenDataMatchingConcepts2012,elmagarmidDuplicateRecordDetection2007,kondaMagellanBuildingEntity2016}.
The two main drawbacks of this approach are the necessary manual tinkering and poor performance on dirty data~\cite{mudgalDeepLearningEntity2018}.
However, in the last few years, the research community has increasingly adopted deep learning~\cite{mudgalDeepLearningEntity2018,ebraheemDistributedRepresentationsTuples2018,nieDeepSequencetoSequenceEntity2019,zhaoAutoEMEndtoendFuzzy2019,brunnerEntityMatchingTransformer2020,liDeepEntityMatching2020}.
While early work focused on custom architectures and trained models from scratch,
the current state of the art focuses on fine-tuning large natural language models such as BERT~\cite{devlinBERTPretrainingDeep2019},
which offers higher accuracy and decreases the need for training examples~\cite{liDeepEntityMatching2020}.
We refer to \cite{barlaugNeuralNetworksEntity2021} for an extensive survey on deep learning for EM.

\textit{Explainable AI}.
There are many ways to explain machine learning models.
Generally, explanations can be either global or local~\cite{duTechniquesInterpretableMachine2019}, 
in other words, explaining the model's behavior as a whole or explaining a single prediction.
Furthermore, we often distinguish between intrinsically interpretable models and post hoc interpretation methods~\cite{molnarInterpretableMachineLearning2019} (which can be model-agnostic or not).
The former are models that are interpretable on their own, like linear regression or decision trees,
while the latter are methods for explaining black-box models.
We refer the reader to one of many extensive sources on the topic~\cite{adadiPeekingBlackBoxSurvey2018,duTechniquesInterpretableMachine2019,gilpinExplainingExplanationsOverview2018,guidottiSurveyMethodsExplaining2019,molnarInterpretableMachineLearning2019}.

A particularly prominent group of approaches are local post hoc attribution methods (e.g., \cite{ribeiroWhyShouldTrust2016,lundbergUnifiedApproachInterpreting2017,sundararajanAxiomaticAttributionDeep2017}),
which aim to explain a prediction by communicating to what degree different parts of the input are to be attributed for the prediction.
LIME~\cite{ribeiroWhyShouldTrust2016} is one of the most prominent among such methods.
It is a model-agnostic method,
and works by randomly removing features of an input example and training a (usually linear) interpretable surrogate model to predict the model's output for these perturbations.
Among other popular local post hoc attribution methods are the game-theoretic-based SHAP~\cite{lundbergUnifiedApproachInterpreting2017} and gradient-based methods (e.g., \cite{sundararajanAxiomaticAttributionDeep2017}).

\textit{Explainable EM}.
The use of explainability techniques for machine learning-based entity matching is still a young subject,
and there has only been a limited amount of previous work.
However, we note that rule-based methods have historically been used to make systems that can be interpreted by experts~\cite{elmagarmidDuplicateRecordDetection2007},
and they represent an alternative way to make explainable matchers~\cite{qianSystemERHumanintheloopSystem2019}.

The authors in \cite{ebaidEXPLAINEREntityResolution2019} demonstrate ExplainER,
a tool for exploring explainable entity matching that provides multiple prominent explainability techniques such as LIME and association rules,
while \cite{thirumuruganathanExplainingEntityResolution2019} discuss challenges and research opportunities.
Further, there have been two significant adaptations of LIME for entity matching,
which we will now describe and contrast to our work.

Mojito~\cite{diciccoInterpretingDeepLearning2019} introduces two versions of LIME for entity matching:
LIME\_DROP and LIME\_COPY.
The former is a straightforward application of LIME similar to how the original authors do text classification using token level feature granularity\footnote{Tokens are typically words or singular values, and can be assumed to be for datasets and experiments in this paper, but does not necessarily have to be for the described methods.},
while in the latter they use attribute level representation and perturb by copying the entire attribute value to the corresponding attribute in the other record instead of removing tokens.
LIME\_COPY is an elegant way to address challenge (\ref{challenge-non-matches}), but leaves more to be wanted.
Firstly, attribute level granularity is too coarse-grained for most cases with longer textual attributes (the extreme case being a single textual attribute).
Secondly, since it is separate from LIME\_DROP, it requires the user to interpret two different kinds of explanations.
Finally, it relies on a matched schema with one-to-one attribute correspondence.

Recently, Landmark~\cite{baraldiUsingLandmarksExplaining2021} was proposed as a two-part improvement over Mojito.
Firstly, it makes two explanations, one per record, and avoids perturbing both records simultaneously,
which effectively solves challenge (1).
Secondly, for record pairs labeled as non-matches, instead of perturbing by randomly copying entire attributes,
it appends every corresponding attribute value from the other record and performs regular token level exclusion perturbation (a technique named double-entity generation),
effectively combining LIME\_DROP and LIME\_COPY.
The authors demonstrate through experiments that their techniques are indeed effective and that Landmark is a substantial improvement over Mojito.
One limitation of the approach is that tokens from the other record are only ever considered to be appended at the end of the corresponding attribute.
This is unfortunate if the matcher is sensitive to the order of tokens (e.g., many products have the brand name first in the title), the schemas are not matched one-to-one, or the data is dirty.
Similar to Mojito, Landmark also makes two different kinds of explanations.

While making important contributions, neither Mojito nor Landmark addresses all three challenges identified in Section~\ref{sec:introduction}.
Only Landmark tackles challenge (\ref{challenge-interaction}).
Both propose a solution to challenge (\ref{challenge-non-matches}), but with important limitations, as we discussed above.
Neither addresses challenge (\ref{challenge-variation}).
Moreover, they do not provide a unified and coherent way of actually communicating or visualizing an explanation to the end-user in the same way the original authors of LIME do --- something we aim to do.

\section{Preliminaries}
In this section, we first present the problem definition and then introduce LIME and describe how it can be adapted for entity matching.

\subsection{Problem Definition}

\textit{Entity Matching.}
Let $A$ and $B$ be two data sources.
A data source is a collection of records following the same schema (they all have the same attributes),
and a record $r = \{(\alpha_j, v_j)\}_j$ is an ordered set of attribute name-value pairs.
The goal of entity matching is to find all pairs $(a,b) \in A \times B$ such that $a$ and $b$ refers to the same real-world entity.
We call such pairs \textit{matches} and all other pairs for \textit{non-matches}.
Since there is a quadratic number of pairs $O(|A||B|)$, inspecting all pairs in $A \times B$ is usually infeasible.
Therefore, one will normally first perform a recall-focused step called blocking~\cite{papadakisBlockingFilteringTechniques2021} to produce a set of candidates $C \subseteq A \times B$ such that $|C| \ll |A \times B|$ while still containing most matches with high probability.
Then we classify every $(a,b) \in C$ as either match or non-match.
In this paper, we focus on the record pair classification part of entity matching.
Therefore, for our purposes, entity matching is a binary classification problem deciding whether a pair of records $(a, b)$ refer to the same real-world entity or not (match or non-match).

\textit{Local Post Hoc Attribution for Entity Matching}.
The goal of a local post hoc attribution explainability method for entity matching is to explain a single prediction of a record pair classification from an arbitrary matcher by communicating the significance (in some shape or form) of the different parts of the two records to the user.
Formally, let a matcher be a classifier ${{f(x)\colon A \times B \to \mathbb{R}}}$ that accepts a record pair ${x = (a, b)}$ such that ${a \in A \land b \in B}$ and outputs a prediction score between $0$ and $1$.
Note that $f$ is not restricted to supervised machine learning models but can be any procedure capable of classifying record pairs with a confidence score.
A local post hoc attribution explainability method for entity matching provide two things.
First, it provides a procedure $\lambda(f, x)$ that accepts a matcher $f$ and a record pair $x$ and outputs an explanation $e_x$.
Secondly, it provides a framework of interpretation for the explanations.
An explanation $e_x$ attributes different parts of $x$ to the prediction score ${y = f(x)}$ using real-valued attribution scores,
and the explanation is communicated to the user either through numbers directly or some visualization (see for example \cite{ribeiroWhyShouldTrust2016}).
How the attribution scores are to be interpreted and how it should be communicated to the user is up the method.

As stated in Section~\ref{sec:introduction}, evaluation of explainability methods is still an open problem and there are no standards for how to do it in entity matching.
Therefore, as part of our contribution we propose ways to do this for attribution methods.
We refer the reader to Section~\ref{sec:experiments} for more on this.

\subsection{LIME}\label{sec:lime}
\begin{figure*}
    \centering
    \includegraphics{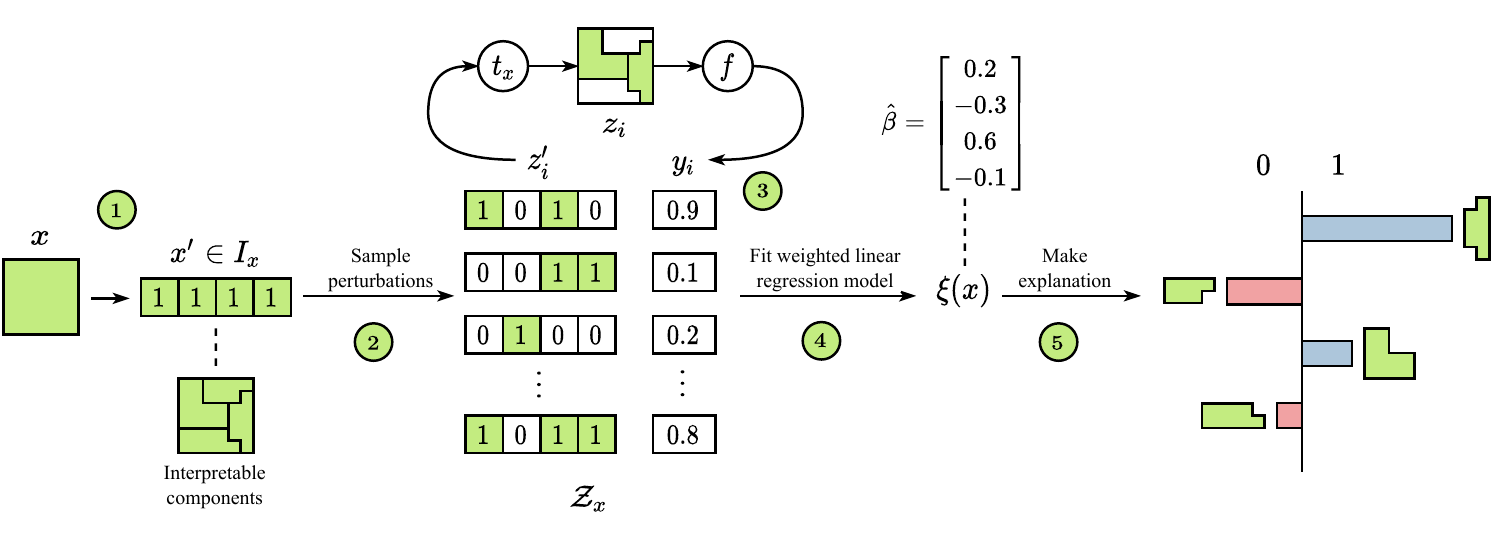}
    \caption{
        Illustration of LIME \cite{ribeiroWhyShouldTrust2016} for a binary classification problem.
        The main steps are:
        1) Split the input $x$ into interpretable components.
        2) Sample neighbors $z'_i$ of the interpretable representation $x' \in I_x$.
        3) Convert each $z'_i$ to a corresponding input domain representation $z = t_x(z'_i)$ and run inference to get $y_i = f(t_x(z'_i))$.
        4) Fit a weighted linear regression model on the neighborhood dataset $\mathcal{Z}_x$ to get $\xi(x)$ --- essentially the regression coefficients $\hat{\beta}$.
        5) Present $\hat{\beta}$ in a user-friendly way relating them to the interpretable components.
    }
    \label{fig:lime-overview}
\end{figure*}
The main idea of LIME \cite{ribeiroWhyShouldTrust2016} is to locally approximate a classifier around one instance with an interpretable model over an interpretable representation in a way that balances faithfulness and complexity, and then use the interpretable model as an explanation.
The intuition is that while our problem is too complex for classical machine learning models that we regard as inherently interpretable (e.g., linear regression or decision trees) to be accurate enough,
it might be possible to faithfully approximate the decision boundary for a black-box model locally around one input instance.
In other words, the authors train an interpretable surrogate model using local data points sampled by perturbing an input instance and use it as an explanation of that particular instance.
And while the input features of a black-box model might be unsuited for human interpretation (e.g., deep learning embeddings or convoluted string metrics),
they define an alternative interpretable representation for the input instance we want to explain and use that to train the interpretable surrogate model.

Formally, let ${f(x)\colon A \times B \to \mathbb{R}}$ be the matcher we want to explain.
Furthermore, for a single instance $x = (a, b)$ that we want to explain, let ${I_x = \{0, 1\}^{d_x}}$ be the interpretable domain and $x' \in I_x$ the interpretable representation of $x$.
Its elements represent the presence (or absence) of what is called \textit{interpretable components} in \cite{ribeiroWhyShouldTrust2016},
essentially non-overlapping parts of the input.
E.g., for text, it could be the presence of different words.
For each $x$ there must exist a function ${t_x(x')\colon I \to A \times\! B}$ that can translate an interpretable representation to the input domain of the classifier $f$.

With the goal of approximating $f$ local to $x$, the authors sample a new dataset ${\mathcal{Z}_x = \{(z', y) \in I \times \mathbb{R}\}}$ where ${y = f(t_x(z'))}$.
Each $z'$ is drawn by setting a uniformly random-sized subset of $x'$ to zero.
Let $G$ be a class of interpretable models over $I_x$.
Furthermore, let $\mathcal{L}(f, g, \pi_x)$ be how unfaithful $g \in G$ is to $f$ in the neighborhood defined by the distance kernel $\pi_x(z')$.
They want to find a $g \in G$ that is as faithful to $f$ as possible, but since many interpretable models can be made more accurate by increasing the complexity, they need to balance the faithfulness with model complexity so $g$ is simple enough to actually be interpretable for humans.
To that end, let $\Omega(g)$ be a measure of complexity for $g$,
and choose the following explanation for $x$:
\begin{equation}
    \xi(x) = \underset{g \in G}{\mathrm{argmin}} \Big[ \mathcal{L}(f,g,\pi_x) + \Omega(g) \Big]
\end{equation}

In their work, the authors only present one concrete instance of their general framework\footnote{We will, as is common in the literature, refer to both the general framework and the described concrete instance as LIME interchangeably.}.
They chose $G$ to be weighted sparse linear regression models and $\mathcal{L}$ to be mean squared error weighted by $\pi_x$ on $\mathcal{Z}_x$.
Furthermore,
$\Omega(g)$ is chosen to be the number of non-zero coefficients of $g$, and the trade-off between $\mathcal{L}$ and $\Omega$ is simplified by constraining $\Omega(g)$ to not be greater than a constant $K$ known to be low enough.
It is now simply a matter of fitting a regularized weighted least squares linear regression model on $\mathcal{Z}_x$.
For a simplified overview of the whole process see Figure~\ref{fig:lime-overview}.

\subsection{LIME for Entity Matching} \label{sec:lime-em}
Before we describe our proposed method in the next section, we will now go through the design decisions done within the LIME framework.
A setup we then build upon and use as a baseline for our proposed method.

Let a record $r = \{(\alpha_j, v_j)\}$ be an ordered set of pairs with attribute name and value.
Inspired by how \cite{ribeiroWhyShouldTrust2016} apply LIME for text classification,
we define the interpretable representation $I_x$ to be the absence of unique tokens in attribute names and values for both records in $x = (a, b)$ (i.e., $t_x(\mathbf{0}) = x$)\footnote{
The reader might also note that the choice of 0/1 semantics are flipped compared to the authors in \cite{ribeiroWhyShouldTrust2016} (see Section~\ref{sec:lime}).
This is simply to be more conceptually similar to our proposed method and is not critical to the approach.
}.
Attribute values that are not strings are treated as single tokens, and their absence is their null/zero value.

\begin{figure}
    \centering
    \includegraphics{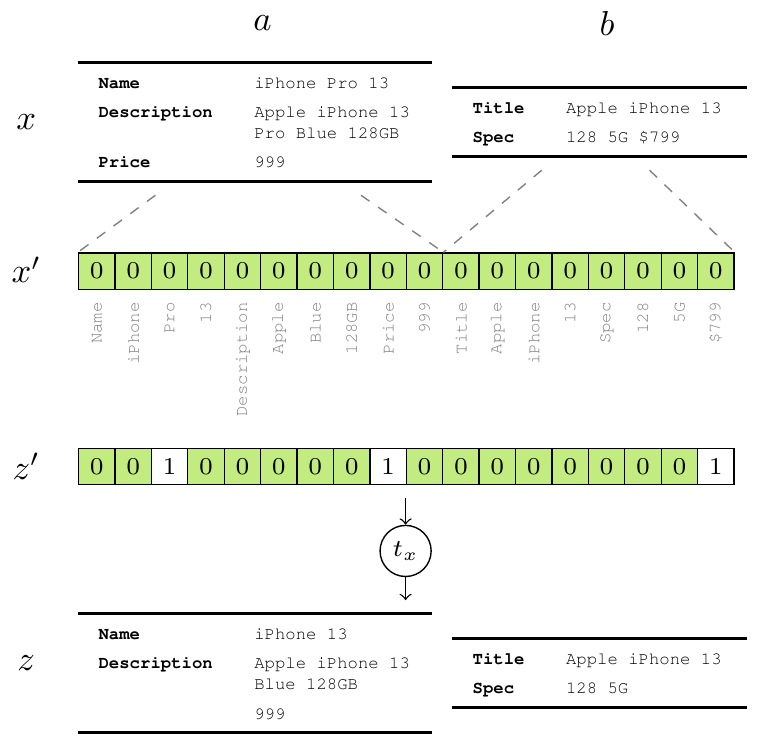}
    \caption{Example record pair $x$, its corresponding interpretable representation $x'$ in LIME, and an example of a perturbed sample.}
    \label{fig:lime-interpretable-representation-example}
\end{figure}

\begin{example}\label{ex:lime}
Figure~\ref{fig:lime-interpretable-representation-example} shows an example record pair $x$ and its corresponding interpretable representation $x'$.
In this example, the records are product descriptions of two similar (but different) phones.
The record $a$ refers to a ``pro'' version of the phone referred to by $b$.
Furthermore, it is uncertain whether the phones have the same color since $b$ does not specify its color while $a$ is blue.
The figure also shows an example of a perturbed interpretable representation $z'$ and how it is translated with $t_x$ into a perturbed record pair $z$.
The token \texttt{Pro} (among others) is removed from record $a$ and it is more likely that a reasonable matcher will consider the record pair $z$ a better match then $x$.
\end{example}

We sample $\mathcal{Z}_x$ by setting random subsets of $x'$ to one, where the size of the subsets are sampled uniformly from the interval $[0, D_{\max}]$,
and we use the neighborhood distance kernel
\begin{equation}
    \pi_x(z') = \exp({-2 \frac{D(x', z')}{D_{\max}}})
\end{equation}
where $D$ is the Hamming distance.
While the original authors simply used $D_{\max} = d_x$ for text classification,
we empirically find this neighborhood too large due to entity matching generally being more sensitive to single tokens compared to standard text classification.
This could, of course, be accounted for by narrowing $\pi_x$, but it is more sample efficient to also reduce the neighborhood we are sampling from.
We use $D_{\max} = max(5, \lfloor \frac{d_x}{5} \rfloor)$,
and let the number of samples $|\mathcal{Z}_x|$ be $\max(500, \min(30 d_x, 3000))$.
From our experience, the results are not sensitive to these parameters.

Finally, we let $G$ be the set of weighted regression models without intercept.
The loss then being
\begin{equation}
    \mathcal{L}(f, g, \pi_x) = \sum_{(z', y) \in \mathcal{Z}_x} \pi_x(z') (y - g(z'))^2
\end{equation}
$\xi(x)$ is found using weighted least squares and forward selection (choosing $K$ coefficients).

\section{Method}
We now describe how we address the three challenges described in Section~\ref{sec:introduction} with three distinct, but coherent, techniques that together form our proposed method: Local explanations for Entities that Match Or Not (LEMON).
As discussed earlier, we use LIME as the basis for our method, but the proposed ideas have wider applicability.
The three following subsections respectively address and propose a solution to the three challenges (1) cross-record interaction effects, (2) non-match explanations, and (3) variation in sensitivity.

\subsection{Dual Explanations}
One shortcoming of LIME, when applied directly to entity matching, is that it does not take into account the inherent duality of the matching.
No distinction is made between the two input records.
This is problematic because our surrogate linear regression model assumes independent features,
but perturbations across two input records will naturally have strong interaction effects.
In essence, the surrogate model $g$ cannot sufficiently capture the behaviour of our matcher $f$ even for small neighborhoods,
which severely hurts the approximation accuracy.

The proposed solution is relatively straightforward but still effective.
Equivalently to \cite{baraldiUsingLandmarksExplaining2021}, we make two explanations, one for each record.
We call such a pair ${(e_x^a, e_x^b)}$ of complementary explanations for dual explanations.
For each explanation, we let $I_x$ represent only the absence of tokens in one record.
In effect, we approximate attributions from only one record at a time while keeping the other constant.
That way, we avoid the strong interaction effects across them.
Intuitively, we explain why record $a$ matches $b$ or not and why record $b$ matches $a$ or not, separately.
The two explanations can still be presented together as one joint explanation to the user.

\subsection{Attribution Potential}\label{sec:potential}
Attribution methods such as the LIME implementation described above tell us which part of the input is the most influential.
This is usually achieved using some kind of exclusion analysis where one looks at the difference between the absence and presence of input features (e.g., \cite{ribeiroWhyShouldTrust2016, sundararajanAxiomaticAttributionDeep2017, lundbergUnifiedApproachInterpreting2017}).
While that might be an effective approach for many machine learning problems,
it is inherently unsuited for entity matching.
The issue lies in explaining non-matches.
Record pairs that a matcher classifies as a match can be explained subtractively because removing or zeroing out essential parts of the records will result in lower matching scores from most well-behaved matchers.
But for record pairs where the matcher is convinced they do not match and provide a near-zero match score,
it is unlikely that removal or zeroing out any part of the records will make a significant difference on the match score.
For example, in the record pair from Example~\ref{ex:lime} a matcher's output might not change significantly by removing \texttt{Blue} from $a$ because it correctly identifies that there is still a lack of matched color information.
Seemingly, nothing influences the match score,
thereby providing no useful signal of the contribution from different features.
Notice that standard gradient-based methods are not able to escape this problem because $f$ will, in these cases, be in a flat area and the gradients be rather uninformative.
Intuitively, we can not explain why two records do not match by what they contain.
A natural solution to this problem is instead to explain by what they do not contain.

\textit{Interpretable Representation}.
Let the interpretable representation $I_x = \{P, A, M\}^{d_x}$ be categorical instead of binary,
where the values represent whether the corresponding token is \textit{Present}, \textit{Absent}, or \textit{Matched}.
Unsurprisingly, $z'_i = A$ means $t_x(z')$ will exclude token $i$ and if $z'_i = P$ it will be kept --- much like before.
On the other hand, if $z'_i = M$, we will copy and inject the token in the other record where it maximizes the match score $f(t_x(z'))$.
For now, let us assume we have an accurate and efficient implementation of $t_x(z')$.

\begin{figure}
    \centering
    \includegraphics{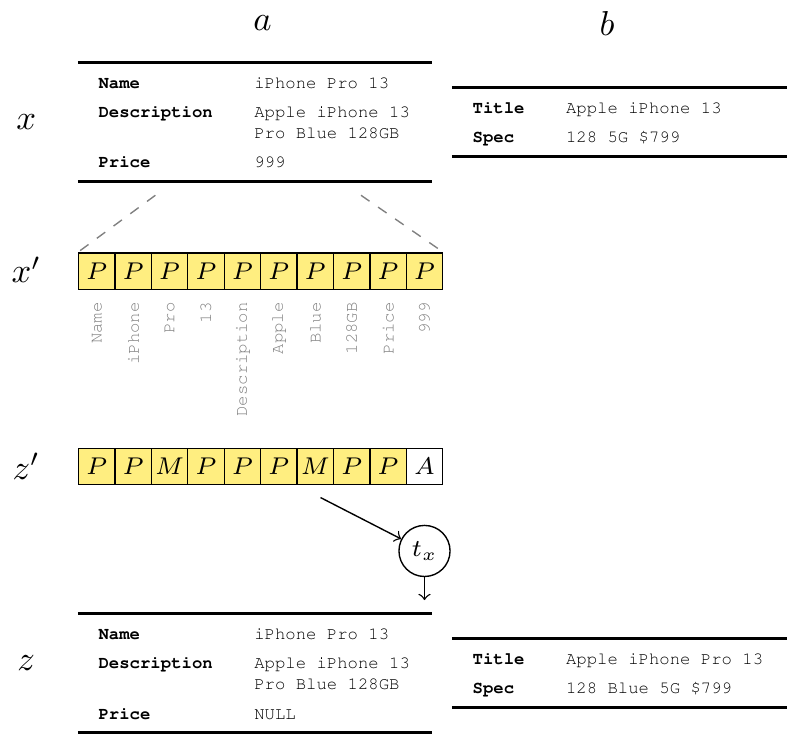}
    \caption{Example record pair $x$, its corresponding interpretable representation $x'$ in LEMON for one of two dual explanations, and an example of a perturbed sample.}
    \label{fig:lemon-interpretable-representation-example}
\end{figure}

\begin{example}
Reusing the record pair $x$ from Example~\ref{ex:lime}, Figure~\ref{fig:lemon-interpretable-representation-example} shows the interpretable representation $x'$ for one of two dual explanations in LEMON (explaining why $a$ matches, or not, $b$).
The representation would, of course, be similar for the other of the two explanations.
In the figure, we also see an example of a perturbed interpretable representation $z'$ and its equivalent record pair $z$ provided by $t_x$.
Notice how both \texttt{Pro} and \texttt{Blue} have been injected into $b$ --- making it more likely to be accepted as a match by a reasonable matcher.
\end{example}

For the linear surrogate model, we dummy code $I_x$, using $P$ as reference value.
When we do forward selection, we select both dummy variables representing a categorical variable at once,
so that we either pick the entire categorical variable or not.
We will then have two estimated coefficients, $\hat{\beta}^A_i$ and $\hat{\beta}^M_i$, for each of the $K$ selected interpretable features.
Finally, we define the attribution for token $i$ to be $w_i = -\hat{\beta}^A_i$,
and the attribution potential to be $p_i = \hat{\beta}^M_i$.
Intuitively, $w_i$ is the contribution of token $i$, and is the same as in LIME,
while $p_i$ can be interpreted as the maximum additional contribution token $i$ could have had if the other record matched better.
Note that it is important to model this new attribution potential through a categorical variable instead of simply adding another binary variable to $I_x$,
because exclusion and injection perturbations of the same token strongly interact with each other and should be mutually exclusive.

\textit{Approximating \bm{$t_x$}}.
In contrast to plain LIME as described in Section~\ref{sec:lime-em},
$t_x$ is now less straightforward to compute.
The difficulty lies in where to inject tokens $i$ for which $z'_i = M$ to maximize $f(t_x(z'_i))$.
Since our method is model-agnostic, the best we can do is try all possible injections.
That would be computationally prohibitive, not only because of the high number of possible injection targets but also because of the exponential growth of combinations when multiple tokens should be injected.
Instead, we can approximate it by sampling $L$ combinations of injection targets and picking the one that gives the highest match score.

The possible injection targets for a token in an attribute value are anywhere in the string attributes of the other record, but without splitting tokens in the target attribute value.
If the token is a non-string value, it can also overwrite attribute values of the same type --- e.g., a number attribute can replace a number attribute in the other record.
Tokens from attribute names can only be injected to attribute names.
When we sample injection targets, we first pick a target attribute uniformly at random and then a random position within that attribute.
In addition, we employ a heuristic to incorporate information about matched schemas if available.
In cases where the schemas are matched, we boost the probability of choosing the corresponding attribute as the target to 50\%.
This makes efficient use of prior knowledge about the schemas while still preserving robustness to dirty data.
To pick the sample size $L$, we first sum the possible injection targets per token to be injected.
We cap the number of targets to three per attribute and ten in total and then let $L$ be the maximum of all tokens to be injected.

\textit{Neighborhood sampling}.
We sample the neighborhood $\mathcal{Z}_x$ much like before.
Now, $x'$ will be a vector of only $P$, and we let $z'_i = A$ instead of $1$ for random subsets.
But we additionally set random subsets of elements to $M$.
The subset size is chosen uniformly at random from $[0, \max(3, \lfloor d_x / 3 \rfloor)]$,
except with a 50\% chance of picking $0$.
The reason we sample $M$ less than $A$ is that injections tend to have more dramatic effects on the match score than exclusions,
and so we consider them to be larger perturbations and want to avoid drowning the exclusion effects.

\subsection{Counterfactual Granularity} \label{sec:counterfactual-granularity}
Depending on the dataset and matcher,
influencing the match score significantly can sometimes require perturbing large parts of the input records.
This is especially true for datasets where records contain many high-quality pieces of information because it provides the matcher with multiple redundant strong signals about whether they match or not.
An example is the iTunes-Amazon dataset, where attributes such as song name, artist name, album name, and more might all agree or disagree at the same time.
The problem is that we want to pick out $K$ important features for the user to focus on,
but in such cases, no single token is likely to be significantly important.
While we could use attribute-level features, that would be unnecessarily coarse-grained for many cases.
Instead, we propose an adaptive strategy where we automatically choose an appropriate explanation granularity.
The idea is to exponentially decrease the granularity of the interpretable features until the attributions and attribution potentials are large enough in magnitude to explain the decision boundary.

\newcommand{\inc}{\mathit{inc}}
\newcommand{\dec}{\mathit{dec}}
\newcommand{\INC}{\mathit{INC}}
\newcommand{\DEC}{\mathit{DEC}}
\newcommand{\CFS}{\mathit{CFS}}

Let $e_x = \{(w_i, p_i)\}_{i \in E_K}$ be the $K$ pairs of attributions and attribution potentials for an explanation $\xi(x)$,
where $E_K$ is the $K$ interpretable features chosen to be used in the regularized linear surrogate model.
Further,
let $\widehat{\inc}_i = \max(-w_i, p_i)$ and $\widehat{\dec}_i = w_i$.
In other words, this is how much perturbation of token $i$ could increase or decrease the match score according to $w_i$ and $p_i$ if you removed the token or injected the token in the other record.
Then, to represent greedy actions increasing the match score,
let $\INC$ be a vector of the elements in $E_K$ with positive $\widehat{\inc}_i$ and sorted by $\widehat{\inc}_i$ in descending order, and similarly for $\DEC$.
We define the predicted counterfactual strength of $k$ steps to be
\begin{equation}
    \widehat{\CFS}^k(e_x) = 
    \begin{cases}
        p - \big[f(x) - \sum\limits_{s=1}^{s=k} \DEC_s \big], & \text{$f(x) > p$} \\[1ex]
        \big[ f(x) + \sum\limits_{s=1}^{s=k} \INC_s \big] - p, & \text{$f(x) \leq p$}
    \end{cases}
\end{equation}
Intuitively, this is to what degree one would assume to surpass the classification threshold $p$ if one performs $k$ greedy actions to change the matcher prediction.
Note that most matchers, as well as those in our experiments, have a classification threshold $p$ of 0.5~\cite{kondaMagellanBuildingEntity2016,liDeepEntityMatching2020}.
Then let the greedy counterfactual strategy $k_g(e_x)$ be the smallest number of steps predicted to be necessary to get a counterfactual strength of at least $\epsilon$:
\begin{equation}
    k_g(e_x) = 
    \begin{cases}
        \min \{k : \widehat{\CFS}^k(e_x) \geq \epsilon \}, & \text{if such $k$ exists} \\
        |\DEC|, & \text{otherwise if $f(x) > p$} \\ 
        |\INC|, & \text{otherwise if $f(x) \leq p$} \\ 
    \end{cases}
\end{equation}
Finally, we define the predicted counterfactual strength of the explanation $\xi(x)$ to simply be ${\widehat{\CFS}(e_x) = \widehat{\CFS}^{k_g(e_x)}(e_x)}$, and the actual counterfactual strength to be:
\begin{equation}
    \CFS(e_x) = 
    \begin{cases}
        p - \Big[f(x) - f\big(t_x(x'_g)\big) \Big], & \text{$f(x) > p$} \\[1ex]
        \Big[ f(x) + f\big(t_x(x'_g)\big) \Big] - p, & \text{$f(x) \leq p$}
    \end{cases}
\end{equation}
where $x'_g$ is the perturbation of the interpretable representation $x'$ corresponding to the greedy counterfactual strategy.

When an explanation's interpretable features represent (up to) $n$ consecutive tokens,
we say that explanation has a granularity of $n$ tokens.
To find our desired granularity, we start with a granularity of one token and then double until we find a granularity that satisfies $\widehat{\CFS}(e_x) \geq \epsilon \land \CFS(e_x) \geq \epsilon$ or no coarser granularity is possible (i.e., all features are whole attributes).
When no granularity satisfies the requirement, we pick the granularity with the highest harmonic mean between $\widehat{\CFS}(e_x)$ and $\CFS(e_x)$, which will favor them to be large and similar.

We call the resulting approach for picking granularity counterfactual granularity.
It will try to find explanations that explain the decision boundary while balancing maximal granularity and faithfulness.
Note that the granularity is chosen independently for each of the two dual explanations.
Decreasing the granularity exponentially avoids a large increase in runtime compared to fixed-step decrease by exploiting the fact that users are likely to be more sensitive to the same constant sized decrease at high granularities than low granularities. I.e., going from a granularity of one token to two tokens feels more substantial than going from a granularity of eight tokens to nine tokens.

\subsection{Summary}

\begin{figure*}
    \centering
    \includegraphics{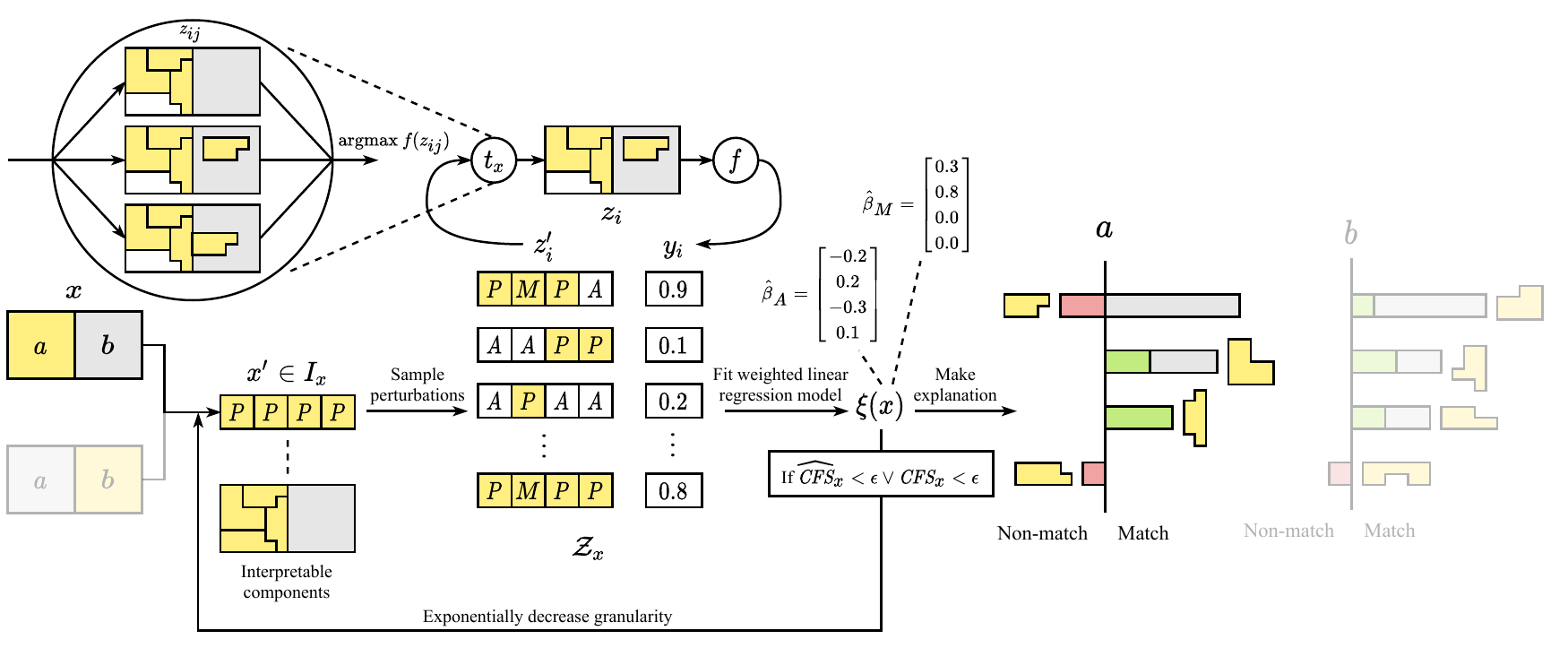}
    \caption{
    Illustration of how LEMON generates an explanation for a record pair $x = (a, b)$.
    The fundamental flow is similar to LIME in Figure~\ref{fig:lime-overview},
    but have been significantly altered and expanded to support dual explanations, attribution potential and counterfactual granularity.
    }
    \label{fig:lemon-overview}
\end{figure*}

\begin{figure*}
    \centering
    \includegraphics[width=1.00\textwidth]{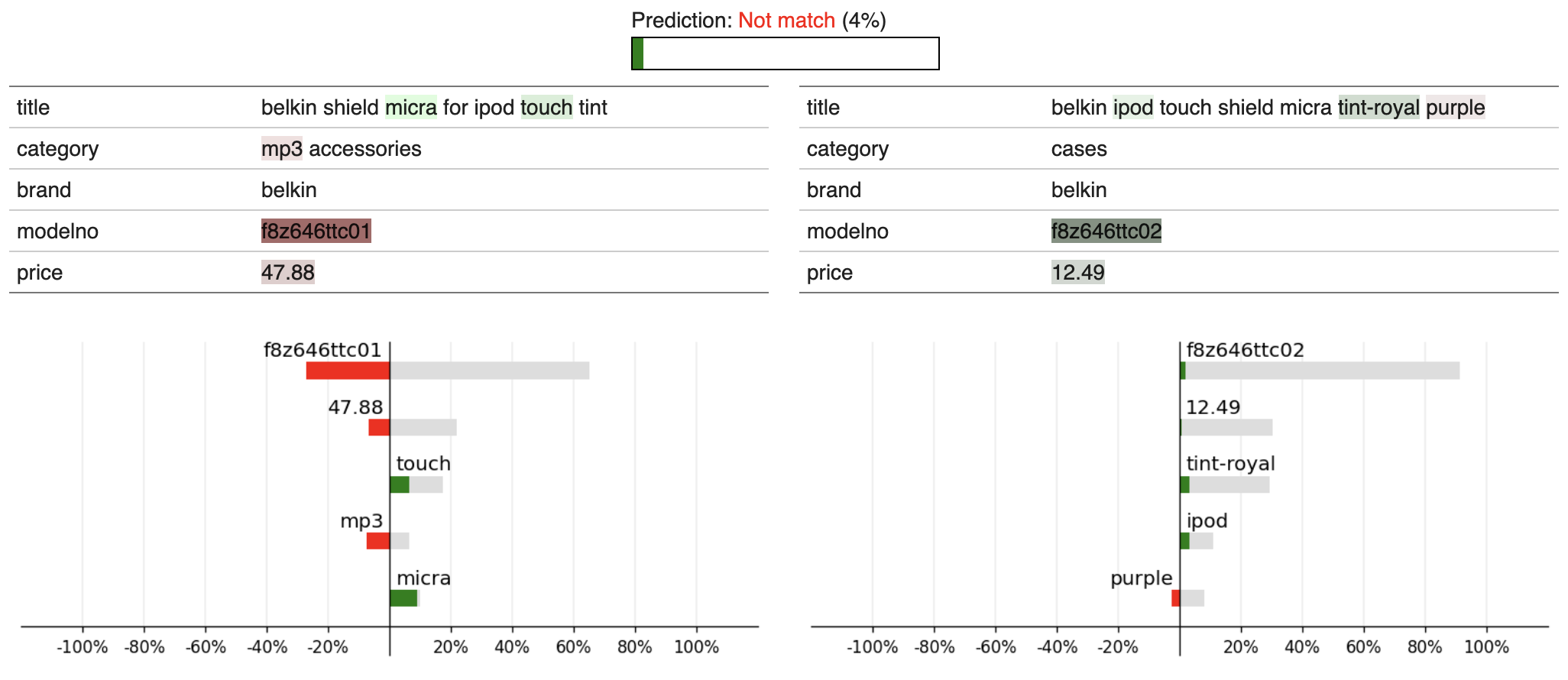}
    \caption{
        Example of how a LEMON explanation can be visualized and what we show users in our user study.
        This particular explanation is for the prediction of the BERT-Mini matcher used in our experiments on the record pair from Table~\ref{fig:record-pair}.
    }
    \label{fig:explanation}
\end{figure*}

\begin{algorithm}
    \caption{LEMON}
    \label{alg:lemon}
    \footnotesize
    \DontPrintSemicolon
    \KwIn{
        Matcher $f$, record pair $(a,b)$, number of features $K$ (default: 5), counterfactual margin $\epsilon$ (default: 0.1), min and max number of samples $S_{\min}$, $S_{\max}$ (default: 500, 3000) \\
    }
    \KwOut{
    Pair of dual explanations $(e_x^a, e_x^b)$.
    Each explanation $e_x = \{ (w_i, p_i) \}_i$ consists of attribution and attribution potential for $K$ chosen interpretable features.
    }
    \vspace{1em}
    \SetKwFunction{FExplain}{Explain}
    \SetKwProg{Fn}{Function}{:}{}
    \Fn{\FExplain{$r$, $o$}}{
    $x \gets (r, o)$\;
    $\overline{\CFS}^* \gets -\infty$\;
    $e_x^* \gets$ null\;
    $n \gets$ 1\;
    $N \gets$ max num. of tokens in any non-empty string in r or 1\;
    \While{$n < 2N$}{
        $
        \begin{aligned}
            x' \gets& \text{ interpretable representation of } x \text{ for } r \text{ with } \\[-0.2em]
            &\text{ granularity of } n \text{ tokens}\;
        \end{aligned}
        $\;
        $\mathcal{Z}_x \gets \{\}$\;
        $S \gets \max(S_{\min}, \min(30 d_x, S_{\max}))$\;
        \For{$i \in \{1, 2, \dots, S\}$}{
            Sample perturbation $z'$ of $x'$\;
            $y \gets f(t_x(z'))$\;
            $\mathcal{Z}_x \gets \mathcal{Z}_x \cup \{(z', y)\}$\;
        }
        $
        \begin{aligned}
        \{(\hat{\beta}^A_i, \hat{\beta}^M_i)\}_i \gets& \text{ linear regression on } \mathcal{Z}_x \text{ weighted by } \pi_x,\\[-0.2em]
        &\text{ selecting only }K \text{ features} \\[-0.2em]
        &\text{ using forward selection}
        \end{aligned}
        $\;
        \vspace{-0.5em}
        $e_x \gets \{(w_i,p_i)\}_i$ calculated from $\{(\hat{\beta}^A_i, \hat{\beta}^M_i)\}_i$\;
        \vspace{0.5em}
        Calculate $\widehat{\CFS}(e_x)$ and $\CFS(e_x)$\;
        \vspace{0.5em}
        \If{$\widehat{\CFS}(e_x) \geq \epsilon \land \CFS(e_x) \geq \epsilon$}{
            \KwRet $e_x$\;
        }
        $\overline{\CFS} \gets \frac{\widehat{\CFS}(e_x) \cdot \CFS(e_x)}{\widehat{\CFS}(e_x) + \CFS(e_x)}$\;
        \vspace{0.5em}
        \If{$\overline{\CFS} > \overline{\CFS}^*$}{
            $\overline{\CFS}^* \gets \overline{\CFS}$\;
            $e_x^* \gets e_x$\;
        }
        $n \gets 2n$\;
    }
    \KwRet $e_x^*$\;
    }
    \vspace{0.5em}
    $e_x^a \gets$ \FExplain{$a$, $b$}\;
    $e_x^b \gets$ \FExplain{$b$, $a$}\;
    \vspace{0.5em}
    \KwRet $(e_x^a, e_x^b)$\;
\end{algorithm}

All three extensions introduced above fit together in our proposed method.
Finally, we choose $K$ to be $5$ and $\epsilon$ to be $0.1$ for all examples.
Figure~\ref{fig:lemon-overview} provides a simplified overview of the steps that make up LEMON,
while Algorithm~\ref{alg:lemon} provides pseudocode.
It is a model-agnostic and schema-flexible method without any hyperparameters that need tuning.
One downside of LEMON is that, due to its extensions, it is more computationally demanding than LIME.
The main reason is the increased number of matcher predictions made to estimate the attribution potential and finding the right granularity.
However, in most cases it is still possible to generate an explanation within a few seconds.

\textit{Complexity.}
To analyze the runtime formally, we focus only on the number of predictions performed using the matcher $f$.
This is reasonable because, for any non-trivial matcher, the runtime is completely dominated by the runtime of the matcher.
Let $F$ be the upper bound on the runtime of $f$ for all possible perturbations of $x$.
Furthermore, let $N$ be the max number of tokens in any non-empty string in $x$ or 1, $S$ be the number of samples $|\mathcal{Z}_x|$ at 1 token granularity, and $L$ be the max number of attribution potential samples for all perturbations of $x$.
The time complexity of LEMON is then $O(F S L \log N)$.
Technically, since $S$ and $L$ are bounded by constants, $O(F \log N)$ would also be accurate, but these factors are essential to understand the difference from similar methods.
LIME~\cite{ribeiroWhyShouldTrust2016}, SHAP~\cite{lundbergUnifiedApproachInterpreting2017}, and Landmark~\cite{baraldiUsingLandmarksExplaining2021} (see Section~\ref{sec:related-work}) are all $O(F S)$, while gradient-based methods are typically $O(F)$.
Ignoring that different methods have different strategies for choosing $|Z_x|$,
the reason for LEMON's increased runtime compared to $LIME$ is the additional factor $L \log N$.
Since $L$ is low and bounded we still get feasible runtime in practice.
See Section~\ref{sec:runtime} for an empirical evaluation.
Note that an analysis of space complexity is less interesting since any non-trivial matcher and dataset will dominate the space requirements compared to the explanation method itself.

\textit{Explanations.}
One key advantage of LEMON over previous work is that it provides one type of explanation for all record pairs, whether the records match or not,
with a clear and easy way to interpret and visualize.
The attributions $w_i$ are equivalent to those in LIME and can be interpreted in the same way.
Its interpretation is to what degree interpretable feature $i$ (some part of a record) contributes to the match score.
If the corresponding part of the record is removed, we expect the match score to decrease by approximately $w_i$.
For the same interpretable feature $i$, the interpretation of the attribution potential $p_i$ is how much higher the attribution $w_i$ could be if the other record matched the feature better.

While the explanation can be visualized in many ways,
we propose a straightforward extension of the visualization proposed by the original LIME authors.
Figure~\ref{fig:explanation} shows an example.
In addition to plotting a colored bar for each $w_i$, we also plot gray bars from $w_i$ to $w_i + p_i$,
outlining feature $i$'s potential attribution.

\section{Experimental Setup}\label{sec:experimental-setup}
\subsection{Datasets}
\begin{table}[]
    \centering
    \caption{
    The Public DeepMatcher~\cite{mudgalDeepLearningEntity2018} Benchmark Dataset and $F_1$ Score for the Magellan and BERT-Mini Matchers Used in the Experiments.
    }
    \scriptsize
    \begin{tabular}{llrrrcc}
    \toprule
        \multirow{2}[1]{*}{\textbf{Type}} & \multirow{2}[1]{*}{\textbf{Name}} & \multirow{2}[1]{*}{\textbf{\#Cand.}} & \multirow{2}[1]{*}{\textbf{\#Matches}} & \multicolumn{2}{c}{\textbf{Matcher $F_1$}} \\
        \cmidrule{5-6}
        & & & & MG & BM \\
    \midrule
        \multirow{7}{*}{Structured}
        & Amazon-Google  & 11 460 & 1 167 & 0.52 & 0.67 \\
        & Beer           & 450    & 68    & 0.85 & 0.76 \\
        & DBLP-ACM       & 12 363 & 2 220 & 0.99 & 0.98 \\
        & DBLP-Scholar   & 28 707 & 5 347 & 0.94 & 0.93 \\
        & Fodors-Zagats  & 946    & 110   & 1.00 & 0.95 \\
        & iTunes-Amazon  & 539    & 132   & 0.90 & 0.93 \\
        & Walmart-Amazon & 10 242 & 962   & 0.66 & 0.80 \\
        \midrule
        \multirow{4}{*}{Dirty}
        & DBLP-ACM       & 12 636 & 2 220 & 0.91 & 0.97 \\
        & DBLP-Scholar   & 28 707 & 5 347 & 0.83 & 0.94 \\
        & iTunes-Amazon  & 539    & 132   & 0.53 & 0.90 \\
        & Walmart-Amazon & 10 242 & 962   & 0.41 & 0.79 \\
        \midrule
        \multirow{2}{*}{Textual}
        & Abt-Buy & 9 575   & 1 028  & 0.51 & 0.81 \\
        & Company & 112 632 & 28 200 & 0.57 & 0.90 \\
    \bottomrule
    \end{tabular}
    \label{tab:datasets}
\end{table}

All experiments are carried out on the 13 public datasets used in the evaluation of DeepMatcher~\cite{mudgalDeepLearningEntity2018} --- originally from \cite{kopckeEvaluationEntityResolution2010} and \cite{dasMagellanDataRepository2015}.
Table~\ref{tab:datasets} lists them together with their number of candidates and number of matches.
The datasets are divided into three types: structured, dirty, and textual.
Structured datasets have nicely separated attributes.
Dirty datasets are created from their structured counterpart by randomly injecting other attributes into the title attribute~\cite{mudgalDeepLearningEntity2018},
and textual datasets generally consist of long textual attributes containing multiple pieces of information.
For the company dataset, we truncate each record to max 256 space-separated words.

When we take a closer look at properties of the different explainability methods and the studied behavior is similar across all datasets we sometimes report only for a subset of the datasets or a single dataset (Abt-Buy) due to space restrictions.
For those experiments one can assume the general behavior is similar on the other datasets.

\subsection{Matchers}\label{sec:matchers}

To show that our proposed method is versatile and model-agnostic, we perform our experiments for each dataset on both a matcher that uses classical machine learning with string metrics as features and on a deep learning based matcher.
See Table~\ref{tab:datasets} for their $F_1$ score on the benchmark datasets.

\textit{Magellan}.
For the classical approach, we train a Magellan~\cite{kondaMagellanBuildingEntity2016} random forest matcher.
We use the automatically suggested similarity features and the default random forest settings provided by the library.
Furthermore, we do not downsample and train on the entire training dataset.

\textit{BERT-Mini}.
For the deep learning approach, we train a baseline Ditto~\cite{liDeepEntityMatching2020} matcher using BERT-Mini~\cite{turcWellReadStudentsLearn2019}.
While not quite achieving state-of-the-art accuracy,
BERT-Mini provides a decent accuracy vs. cost trade-off while maintaining the main characteristics of state-of-the-art deep learning matchers and still significantly outperforming the classical matcher on dirty and textual data\footnote{Since we perform an extensive set of experiments we want to be mindful of our usage of computational resources --- both to reduce the energy footprint and keep the experiments as accessible as possible. For the purpose of this paper we consider this matcher to be sufficiently representative of state-of-the-art matchers. See Appendix~\ref{sec:other-matchers} for results on the main experiments for a RoBERTa-based~\cite{liuRoBERTaRobustlyOptimized2019} DITTO matcher and DeepMatcher~\cite{mudgalDeepLearningEntity2018}}.
We use batch size 32, linearly decreasing learning rate from $3 \cdot 10^{-5}$ with 50 warmup steps, 16-bit precision optimization,
and 1, 3, 5, 10, or 20 epochs depending on the dataset size.
The final model is the one from the epoch with the highest F1 score on the validation dataset.

\subsection{Baselines}
We now introduce the baselines we use for comparison.
For a fair comparison, we adopt dual explanations for all of them.

\textit{LIME}.
Since our work can be seen as a continuation of LIME~\cite{ribeiroWhyShouldTrust2016},
it is a natural baseline.
We use LIME as described in Section~\ref{sec:lime-em}.

\textit{Landmark}.
This is the most relevant work to ours (see Section~\ref{sec:related-work}).
We use the source code provided by the authors\footnote{\url{https://github.com/softlab-unimore/landmark}} with default settings.

\textit{SHAP}.
Another popular approach for producing input attributions is SHAP~\cite{lundbergUnifiedApproachInterpreting2017}.
It is based on the game-theoretic Shapley values~\cite{lipovetskyAnalysisRegressionGame2001,strumbeljExplainingPredictionModels2014} and provides several methods for different types of models.
For a fair comparison, we use their model-agnostic method, Kernel SHAP, which can be interpreted as using LIME to approximate Shapley values.
Note that Kernel SHAP does not limit $K$ and sets $\Omega(g) = 0$.
We use default settings from the SHAP library.

\textit{Integrated gradients}.
Our proposed method is a perturbation-based attribution method.
To compare against a gradient-based method, we use integrated gradients~\cite{sundararajanAxiomaticAttributionDeep2017} as a baseline.
This method is not truly model-agnostic as it requires gradients, so we can only apply it to our deep learning matcher.

Integrated gradients explain input $x$ in reference to some baseline input $x^*$ (some neutral input that gives a score close to zero).
Let $x$ be the embedding vector of a record pair.
As the authors suggest for textual input, we let $x^*$ be the zero embedding.
The attribution for the $i$th element of $x$ is then defined to be
\begin{equation}
    IG_i(x_i) = (x_i - x_i^*) \times \int_{\alpha=0}^1 \frac{\partial f(x^* + \alpha \times (x - x^*))}{\partial x_i} d\alpha
\end{equation}
The integral is approximated by averaging the gradient of evenly spaced points from $x^*$ to $x$.
We use 50 points in our experiments.
The raw attributions are for single elements of the embedded input,
by no means interpretable for humans,
so it is common to sum them for each embedding.
To get attributions on the same representation level as our method, we combine attributions of Bert subword embeddings into whole words.

\section{Experiments}\label{sec:experiments}

We will now go through several experiments to evaluate LEMON and compare it to other methods.
When we evaluate post hoc explainability,
it is important to remember that we do not wish to measure the performance of the matchers,
but rather what the explainability method can tell us about the matchers.
Explanations should not be judged disconnected from the matcher on whether they provide the same rationale as users but to what degree they reflect the actual (correct or wrong) behavior of the matchers and to what degree they are effective at communicating this to users.
Note that some experiments report only results for one or a few datasets when the results tend be similar,
due to space constraints.
Please see Appendix~\ref{sec:extensive-results} for extensive results.

\subsection{Counterfactual Interpretation}\label{sec:counterfactual-interpretation}

\begin{table*}[h]
    \centering
    \caption{
        Counterfactual $F_1$ Score for All the Evaluated Explainability Methods Across All Datasets and the Two Matchers.
    }
    \scriptsize
    \begin{tabular}{cclcccccccccccccc}
        \toprule
        \multirow{3}{*}[-0.4em]{\textbf{Model}} & \multirow{3}{*}[-0.4em]{\textbf{Type}} & \multirow{3}{*}[-0.4em]{\textbf{Method}} & \multicolumn{13}{c}{\textbf{Dataset}} & \\
        & & & \multicolumn{7}{c}{Structured} & \multicolumn{4}{c}{Dirty} & \multicolumn{2}{c}{Textual} \\
        \cmidrule(lr){4-10} \cmidrule(lr){11-14} \cmidrule(lr){15-16}
        & & & AG & B & DA & DG & FZ & IA & WA & DA & DG & IA & WA & AB & C & Mean \\
        \midrule
         \multirow{16}{*}{Magellan} & \multirow{7}{*}{Match}
          & LIME             &          0.96 &          0.83 & \textbf{1.00} &          0.87 &          0.77 &          0.98 &          0.82 &          0.50 &          0.79 &          0.90 &          0.86 &          0.95 &          0.47 &          0.82 \\
        & & SHAP             &          0.95 & \textbf{1.00} & \textbf{1.00} & \textbf{1.00} &          0.95 & \textbf{1.00} &          0.83 &          0.65 &          0.96 &          0.68 &          0.81 &          0.98 &          0.68 &          0.88 \\
        & & SHAP (w/ CFG)    & \textbf{1.00} & \textbf{1.00} & \textbf{1.00} & \textbf{1.00} & \textbf{1.00} & \textbf{1.00} &          0.98 &          0.99 & \textbf{1.00} &          0.91 & \textbf{0.99} & \textbf{0.99} & \textbf{0.87} & \textbf{0.98} \\
        & & Landmark         &          0.92 &          0.89 & \textbf{1.00} &          0.96 &          0.73 &          0.87 &          0.95 &          0.75 &          0.74 &          0.88 &          0.90 &          0.95 &          0.28 &          0.83 \\
        & & LEMON (w/o DE)   &          0.98 &          0.91 & \textbf{1.00} &          0.97 &          0.95 & \textbf{1.00} &          0.96 &          0.83 &          0.93 &          0.93 &          0.95 &          0.98 &          0.49 &          0.91 \\
        & & LEMON (w/o AP)   & \textbf{1.00} & \textbf{1.00} & \textbf{1.00} & \textbf{1.00} & \textbf{1.00} & \textbf{1.00} & \textbf{1.00} & \textbf{1.00} & \textbf{1.00} & \textbf{1.00} & \textbf{0.99} &          0.98 &          0.83 & \textbf{0.98} \\
        & & LEMON (w/o CFG)  &          0.96 &          0.89 & \textbf{1.00} &          0.88 &          0.91 &          0.98 &          0.81 &          0.52 &          0.81 &          0.85 &          0.83 &          0.92 &          0.41 &          0.83 \\
        & & LEMON            & \textbf{1.00} & \textbf{1.00} & \textbf{1.00} & \textbf{1.00} & \textbf{1.00} & \textbf{1.00} & \textbf{1.00} & \textbf{1.00} & \textbf{1.00} & \textbf{1.00} &          0.98 & \textbf{0.99} &          0.80 & \textbf{0.98} \\
        \cmidrule(l){2-17}
        & \multirow{8}{*}{Non-match}
          & LIME             &          0.02 &          0.11 &          0.02 &          0.01 &          0.02 &          0.14 &          0.03 &          0.09 &          0.10 &          0.17 &          0.10 &          0.13 &          0.11 &          0.08 \\
        & & SHAP             &          0.00 &          0.03 &          0.00 &          0.01 &          0.00 &          0.05 &          0.02 &          0.07 &          0.06 &          0.10 &          0.12 &          0.06 &          0.13 &          0.05 \\
        & & SHAP (w/ CFG)    &          0.02 &          0.22 &          0.01 &          0.02 &          0.00 &          0.05 &          0.02 &          0.09 &          0.09 &          0.23 &          0.21 &          0.08 & \textbf{1.00} &          0.16 \\
        & & Landmark         &          0.14 & \textbf{0.84} & \textbf{0.14} &          0.20 &          0.23 &          0.21 & \textbf{0.93} &          0.04 &          0.43 &          0.39 &          0.03 &          0.79 &          0.09 &          0.34 \\
        & & LEMON (w/o DE)   &          0.59 &          0.78 &          0.06 &          0.37 &          0.65 &          0.64 &          0.82 &          0.64 &          0.69 & \textbf{0.87} &          0.82 & \textbf{0.88} &          0.96 &          0.68 \\
        & & LEMON (w/o AP)   &          0.04 &          0.42 &          0.02 &          0.03 &          0.02 &          0.14 &          0.05 &          0.17 &          0.21 &          0.26 &          0.36 &          0.17 &          0.68 &          0.20 \\
        & & LEMON (w/o CFG)  &          0.40 &          0.46 &          0.08 &          0.13 &          0.03 &          0.24 &          0.73 &          0.23 &          0.49 &          0.69 &          0.63 &          0.78 &          0.13 &          0.38 \\
        & & LEMON            & \textbf{0.71} &          0.50 &          0.12 & \textbf{0.54} & \textbf{0.98} & \textbf{0.77} &          0.76 & \textbf{0.75} & \textbf{0.78} & \textbf{0.87} & \textbf{0.87} &          0.87 &          0.96 & \textbf{0.73} \\
        \midrule
        \multirow{20}{*}{BERT-Mini} & \multirow{10}{*}{Match}
          & LIME             &          0.95 &          0.65 &          0.97 &          0.85 &          0.93 &          0.65 &          0.81 &          0.97 &          0.69 &          0.62 &          0.80 &          0.78 &          0.18 &          0.76 \\
        & & SHAP             &          0.90 &          0.81 &          0.79 &          0.65 &          0.91 &          0.69 &          0.71 &          0.79 &          0.62 &          0.63 &          0.75 &          0.77 &          0.25 &          0.71 \\
        & & SHAP (w/ CFG)    &          0.95 &          0.96 &          0.92 &          0.98 &          0.86 &          0.89 &          0.97 &          0.98 &          0.99 & \textbf{1.00} &          0.99 & \textbf{1.00} & \textbf{0.39} &          0.91 \\
        & & IG               &          0.90 &          0.43 &          0.71 &          0.83 &          0.50 &          0.70 &          0.67 &          0.69 &          0.89 &          0.81 &          0.68 &          0.79 &          0.33 &          0.69 \\
        & & IG (w/ CFG)      &          0.88 &          0.52 &          0.66 &          0.94 &          0.68 &          0.70 &          0.77 &          0.86 &          0.95 &          0.94 &          0.87 &          0.93 &          0.28 &          0.77 \\
        & & Landmark         &          0.98 &          0.93 & \textbf{1.00} &          0.94 &          0.86 & \textbf{0.94} &          0.84 &          0.99 &          0.83 &          0.90 &          0.88 &          0.83 &          0.08 &          0.85 \\
        & & LEMON (w/o DE)   &          0.99 &          0.69 & \textbf{1.00} &          0.97 &          0.98 &          0.88 &          0.83 & \textbf{1.00} &          0.94 &          0.89 &          0.84 &          0.86 &          0.25 &          0.85 \\
        & & LEMON (w/o AP)   & \textbf{1.00} & \textbf{1.00} & \textbf{1.00} & \textbf{1.00} & \textbf{1.00} & \textbf{0.94} & \textbf{0.99} & \textbf{1.00} & \textbf{1.00} & \textbf{1.00} & \textbf{1.00} &          0.98 & \textbf{0.39} & \textbf{0.95} \\
        & & LEMON (w/o CFG)  &          0.95 &          0.65 &          0.98 &          0.86 &          0.98 &          0.58 &          0.81 &          0.99 &          0.70 &          0.59 &          0.81 &          0.79 &          0.15 &          0.76 \\
        & & LEMON            & \textbf{1.00} & \textbf{1.00} & \textbf{1.00} & \textbf{1.00} & \textbf{1.00} & \textbf{0.94} &          0.98 & \textbf{1.00} &          0.99 & \textbf{1.00} & \textbf{1.00} &          0.97 &          0.37 &          0.94 \\
        \cmidrule(l){2-17}
        & \multirow{10}{*}{Non-match}
          & LIME             &          0.13 &          0.06 &          0.01 &          0.04 &          0.04 &          0.13 &          0.08 &          0.02 &          0.04 &          0.23 &          0.07 &          0.05 &          0.03 &          0.07 \\
        & & SHAP             &          0.14 &          0.16 &          0.01 &          0.04 &          0.01 &          0.29 &          0.08 &          0.02 &          0.05 &          0.33 &          0.14 &          0.15 &          0.18 &          0.12 \\
        & & SHAP (w/ CFG)    &          0.14 &          0.26 &          0.02 &          0.04 &          0.01 &          0.34 &          0.11 &          0.02 &          0.05 &          0.49 &          0.17 &          0.37 &          0.26 &          0.18 \\
        & & IG               &          0.07 &          0.00 &          0.00 &          0.03 &          0.01 &          0.00 &          0.02 &          0.01 &          0.01 &          0.00 &          0.03 &          0.07 &          0.02 &          0.02 \\
        & & IG (w/ CFG)      &          0.08 &          0.03 &          0.00 &          0.04 &          0.01 &          0.00 &          0.03 &          0.02 &          0.02 &          0.02 &          0.06 &          0.08 &          0.03 &          0.03 \\
        & & Landmark         &          0.40 &          0.70 &          0.05 &          0.17 &          0.50 &          0.63 &          0.64 &          0.07 &          0.35 &          0.50 &          0.74 &          0.67 &          0.01 &          0.42 \\
        & & LEMON (w/o DE)   &          0.75 &          0.93 &          0.55 & \textbf{0.68} & \textbf{0.86} &          0.91 & \textbf{0.93} & \textbf{0.74} &          0.78 & \textbf{0.88} & \textbf{0.97} &          0.93 & \textbf{0.97} & \textbf{0.84} \\
        & & LEMON (w/o AP)   &          0.18 &          0.08 &          0.03 &          0.08 &          0.05 &          0.67 &          0.14 &          0.04 &          0.09 &          0.56 &          0.16 &          0.23 &          0.26 &          0.20 \\
        & & LEMON (w/o CFG)  &          0.50 &          0.92 &          0.02 &          0.19 &          0.85 &          0.94 &          0.89 &          0.04 &          0.18 &          0.76 &          0.95 &          0.90 &          0.96 &          0.62 \\
        & & LEMON            & \textbf{0.81} & \textbf{0.94} & \textbf{0.65} & \textbf{0.68} & \textbf{0.86} & \textbf{0.97} &          0.90 &          0.50 & \textbf{0.79} &          0.87 &          0.95 & \textbf{0.98} & \textbf{0.97} & \textbf{0.84} \\
        \bottomrule
    \end{tabular}
    \label{tab:simulated-cff1}
\end{table*}

Explanations can sometimes provide enough information to the user to understand how the prediction could be different.
The authors of \cite{baraldiUsingLandmarksExplaining2021} call this the ``interest'' of an explanation.
We argue similarly that a useful explanation should implicitly reveal to the user some changes to the records that would flip the prediction outcome.
But we further argue that we should help the user understand a minimal number of such changes necessary,
since that would mean the user has a greater understanding of where the decision boundary is.

To that end, we simulate users being shown an explanation for a record pair and then being asked what they think would be some minimal changes to the records that would flip the matcher's prediction.
The simulated users will greedily try to make the smallest number of perturbations necessary according to the explanation, as described in Section~\ref{sec:counterfactual-granularity}.
Of the two dual explanations, they pick the explanation with the lowest $k_g$ if $\widehat{\CFS}(e_x) \geq \epsilon$ or the one with the highest $\widehat{\CFS}(e_x)$ otherwise.
We extend the same greedy strategy to Landmark explanations but with their corresponding perturbations.
Let the \textit{counterfactual recall} of an attribution method be the fraction of explanations where at least one of the two dual explanations indicate how the matching prediction could be flipped ($\widehat{\CFS}(e_x) \geq \epsilon$),
and the \textit{counterfactual precision} be the fraction of those where the greedy counterfactual strategy is actually successful ($\CFS(e_x) > 0$).
To unify them into a single metric, we report the counterfactual $F_1$ score.
We formalize this in the following definition.
\begin{definition}[Counterfactual Recall, Precision, and $F_1$]
Let $\lambda$ be an entity matching attribution method that outputs dual explanations $(e_x^a, e_x^b)$ and let $C \subseteq A \times B$ be a collection of pairs $(a, b)$ (i.e., a dataset).
The \textit{counterfactual recall} of the method $\lambda$ for the matcher $f$ on the record pair collection $C$ is
\begin{equation*}
    \mathit{CR}(\lambda, f, C) = \mathbb{E}_{(a,b) \sim C} \Big[ \max \big( \widehat{\CFS}(e_x^a), \widehat{\CFS}(e_x^b) \big) \geq \epsilon \Big]
\end{equation*}
where $[\dots]$ are Iverson brackets and we assume ${(e_x^a, e_x^b) = \lambda(f, (a,b))}$.
Furthermore, let the recalled pairs in $C$ be
\begin{equation*}
    C_r = \Big\{(a,b) | (a,b) \in C \land \max \big( \widehat{\CFS}(e_x^a), \widehat{\CFS}(e_x^b) \big) \geq \epsilon \Big\}
\end{equation*}
and let the greedy pick among the dual explanations be
\begin{equation*}
    e_x^g = 
    \begin{cases}
        e_x^a, & \text{if $k_g(e_x^a) < k_g(e_x^b)$} \\
        & \text{or $k_g(e_x^a) = k_g(e_x^b) \land \widehat{\CFS}(e_x^a) \geq \widehat{\CFS}(e_x^b)$} \\[1ex]
        e_x^b, & \text{otherwise} \\ 
    \end{cases}
\end{equation*}
The \textit{counterfactual precision} of the method $\lambda$ for the matcher $f$ on the record pair collection $C$ is
\begin{equation*}
    \mathit{CP}(\lambda, f, C) = \mathbb{E}_{(a,b) \sim C_r} \big[ \CFS(e_x^g) > 0 \big]
\end{equation*}
Lastly, the \textit{counterfactual $F_1$ score} is then simply
\begin{equation*}
    \mathit{CF}_1(\lambda, f, C) = \frac{2 \mathit{CR}(\lambda, f, C) \cdot \mathit{CP}(\lambda, f, C)}{\mathit{CR}(\lambda, f, C) + \mathit{CP}(\lambda, f, C)}
\end{equation*}
\end{definition}

We produce 500 explanations for both predicted matches and non-matches (or all when there are less than 500 available) for each explanation method per dataset\footnote{The same pairs are used for the different explanation methods.}.
Table~\ref{tab:simulated-cff1} shows the counterfactual $F_1$ for the different explainability methods, matchers, and datasets\footnote{See Appendix~\ref{sec:extensive-results} for counterfactual precision and recall numbers.}.

We observe that LEMON performs best overall, with the highest or close to the highest $F_1$ score in most cases.
It significantly outperforms all three baselines, where non-matches, as expected, have the most pronounced difference.
Since all the baselines are fundamentally analyzing the prediction by observing what happens when features are removed,
they suffer from the same issue of explaining non-matches as discussed in Section~\ref{sec:potential}.
Importantly,
the low performance of all the baselines backs up the claim that standard local post hoc attribution methods do not work satisfactorily out of the box for entity matching.
Our proposed method generally outperforms Landmark, with the exception of the three datasets for the Magellan matcher on non-matches.
We note that LEMON has the biggest advantage over Landmark on datasets that typically would require more substantial perturbations to flip the prediction,
such as matches in DBLP-GoogleScholar and Company.
At the same time, it is clear that all methods struggle with non-matches on DBLP-ACM and DBLP-GoogleScholar (and Beer to a certain degree) more than other datasets --- especially for Magellan.
This is mainly because the datasets yield a binary classification problem with large margins for the decision boundary.
The classification problem is too easy and the matchers too certain.
The true matches contain many highly similar attributes, while true non-matches tend to have several significantly dissimilar attributes.
Changing the matcher's prediction from non-match to match is hard because it requires many perturbations across most attributes.
Therefore, the reason Landmark performs better in some cases with non-matches for Magellan is mainly because Landmark does not restrict the number of interpretable features to use in the explanation.
This enables higher counterfactual recall at the expense of more complex and less specific explanations.

\subsection{Explanation Faithfulness}\label{sec:explanation-faithfulness}

\begin{table*}[h]
    \centering
    \caption{
        Perturbation Error $\mathit{PE}$ for All the Evaluated Explainability Methods Across All Datasets and the Two Matchers.
    }
    \scriptsize
    \begin{tabular}{cclcccccccccccccc}
        \toprule
        \multirow{3}{*}[-0.4em]{\textbf{Model}} & \multirow{3}{*}[-0.4em]{\textbf{Type}} & \multirow{3}{*}[-0.4em]{\textbf{Method}} & \multicolumn{13}{c}{\textbf{Dataset}} & \\
        & & & \multicolumn{7}{c}{Structured} & \multicolumn{4}{c}{Dirty} & \multicolumn{2}{c}{Textual} \\
        \cmidrule(lr){4-10} \cmidrule(lr){11-14} \cmidrule(lr){15-16}
        & & & AG & B & DA & DG & FZ & IA & WA & DA & DG & IA & WA & AB & C & Mean \\
        \midrule
         \multirow{16}{*}{Magellan} & \multirow{8}{*}{Match}
          & LIME             &         0.33 &          0.38 &          0.31 &          0.34 &          0.31 & \textbf{0.25} &          0.67 &          0.47 &          0.41 & \textbf{0.35} &          0.46 &          0.33 &          0.60 &          0.40 \\
        & & SHAP             &         1.09 &          1.03 &          0.96 &          1.01 &          0.80 &          0.93 &          1.12 &          1.03 &          1.25 &          2.37 &          2.50 &          1.07 &          7.70 &          1.76 \\
        & & SHAP (w/ CFG)    &         1.03 &          0.79 &          0.96 &          1.00 &          0.79 &          0.95 &          0.31 &          0.64 &          1.17 &          1.24 &          1.28 &          1.04 &          3.58 &          1.14 \\
        & & Landmark         &         0.95 &          0.73 &          0.70 &          1.18 &          0.67 &          0.94 &          0.86 &          7.01 &          2.93 &          5.40 &          1.93 &          1.25 &          3.45 &          2.15 \\
        & & LEMON (w/o DE)   &         0.36 &          0.38 & \textbf{0.30} &          0.36 &          0.33 & \textbf{0.25} &          0.35 &          0.45 &          0.40 &          0.40 &          0.37 &          0.33 &          0.68 &          0.38 \\
        & & LEMON (w/o AP)   &         0.33 &          0.40 &          0.31 & \textbf{0.33} & \textbf{0.28} & \textbf{0.25} & \textbf{0.23} & \textbf{0.42} & \textbf{0.39} &          0.37 & \textbf{0.35} &          0.33 & \textbf{0.53} & \textbf{0.35} \\
        & & LEMON (w/o CFG)  &         0.33 & \textbf{0.36} & \textbf{0.30} &          0.34 &          0.30 &          0.26 &          0.27 &          0.45 &          0.40 & \textbf{0.35} &          0.39 &          0.33 &          0.67 &          0.37 \\
        & & LEMON            &\textbf{0.32} &          0.37 & \textbf{0.30} & \textbf{0.33} & \textbf{0.28} &          0.26 &          0.28 & \textbf{0.42} & \textbf{0.39} & \textbf{0.35} &          0.36 & \textbf{0.32} &          0.54 & \textbf{0.35} \\
        \cmidrule(l){2-17}
        & \multirow{8}{*}{Non-match}
          & LIME             &         0.53 &          0.64 & \textbf{0.42} &          0.47 &          0.58 &          0.34 &          0.45 &          0.56 &          0.60 &          0.46 &          0.59 &          0.46 &          0.79 &          0.53 \\
        & & SHAP             &         0.89 &          0.93 &          1.34 &          1.08 &          1.43 &          1.10 &          1.16 &          1.44 &          1.70 &          2.15 &          1.24 &          1.34 &          6.36 &          1.71 \\
        & & SHAP (w/ CFG)    &         0.62 &          0.52 &          0.79 &          0.80 &          1.12 &          0.38 &          0.64 &          0.83 &          0.79 &          0.99 &          0.68 &          0.77 &          0.82 &          0.75 \\
        & & Landmark         &         0.64 &          0.74 &          0.81 &          0.79 &          0.76 &          1.45 &          0.63 &          0.90 &          1.04 &          5.58 &          3.05 &          1.34 &          2.10 &          1.53 \\
        & & LEMON (w/o DE)   &         0.44 &          0.52 &          0.63 &          0.50 &          0.46 &          0.37 & \textbf{0.38} &          0.59 &          0.51 & \textbf{0.40} &          0.46 & \textbf{0.33} & \textbf{0.47} &          0.47 \\
        & & LEMON (w/o AP)   &         0.46 & \textbf{0.35} &          0.47 & \textbf{0.46} &          0.51 & \textbf{0.21} &          0.40 & \textbf{0.49} &          0.51 &          0.44 & \textbf{0.41} &          0.40 &          0.59 & \textbf{0.44} \\
        & & LEMON (w/o CFG)  &         0.43 &          0.55 &          0.44 &          0.50 & \textbf{0.43} &          0.51 & \textbf{0.38} &          0.58 &          0.52 &          0.44 &          0.48 &          0.35 &          0.78 &          0.49 \\
        & & LEMON            &\textbf{0.42} &          0.59 &          0.64 &          0.53 &          0.45 &          0.43 &          0.40 &          0.54 & \textbf{0.49} &          0.46 &          0.47 &          0.37 & \textbf{0.47} &          0.48 \\
        \midrule
        \multirow{20}{*}{BERT-Mini} & \multirow{10}{*}{Match}
          & LIME             &\textbf{0.32} &          0.47 &          0.39 & \textbf{0.40} & \textbf{0.30} &          0.44 &          0.61 & \textbf{0.38} & \textbf{0.41} &          0.48 &          0.62 &          0.77 &          1.20 &          0.52 \\
        & & SHAP             &         0.67 &          0.50 &          1.48 &          0.87 &          0.65 &          0.61 &          1.14 &          1.37 &          0.85 &          0.81 &          1.08 &          0.82 &          2.18 &          1.00 \\
        & & SHAP (w/ CFG)    &         0.64 &          0.57 &          1.03 &          0.78 &          0.61 &          0.71 &          0.74 &          0.91 &          0.71 &          0.71 &          0.71 &          0.60 &          0.60 &          0.72 \\
        & & IG               &         1.20 &          0.76 &          1.64 &          1.15 &          1.01 &          0.80 &          1.18 &          1.65 &          1.05 &          1.16 &          1.17 &          0.89 &          1.93 &          1.20 \\
        & & IG (w/ CFG)      &         1.16 &          0.72 &          1.45 &          1.06 &          0.87 &          0.93 &          1.11 &          1.28 &          0.92 &          0.89 &          0.95 &          0.80 &          0.84 &          1.00 \\
        & & Landmark         &         0.92 &          1.00 &          1.17 &          0.81 &          0.55 &          0.74 &          0.94 &          1.20 &          0.78 &          1.01 &          1.14 &          0.79 &          0.91 &          0.92 \\
        & & LEMON (w/o DE)   &         0.38 &          0.51 &          0.50 &          0.45 &          0.41 &          0.41 &          0.50 &          0.52 &          0.46 &          0.48 &          0.51 &          0.56 & \textbf{0.40} &          0.47 \\
        & & LEMON (w/o AP)   &\textbf{0.32} & \textbf{0.40} &          0.45 &          0.43 &          0.33 & \textbf{0.31} & \textbf{0.45} &          0.40 & \textbf{0.41} & \textbf{0.30} & \textbf{0.42} &          0.49 &          0.61 & \textbf{0.41} \\
        & & LEMON (w/o CFG)  &         0.33 &          0.42 & \textbf{0.38} & \textbf{0.40} &          0.31 &          0.43 &          0.51 & \textbf{0.38} & \textbf{0.41} &          0.44 &          0.54 &          0.60 &          0.68 &          0.45 \\
        & & LEMON            &         0.33 &          0.45 &          0.45 &          0.44 &          0.40 &          0.43 &          0.49 &          0.42 &          0.44 &          0.39 &          0.46 & \textbf{0.47} &          0.53 &          0.44 \\
        \cmidrule(l){2-17}
        & \multirow{10}{*}{Non-match}
          & LIME             &          0.50 &          0.51 &          0.68 & \textbf{0.46} &          0.61 &          0.50 &          0.61 &          0.58 &          0.52 &          0.53 &          0.67 &          0.63 &          0.61 &          0.57 \\
        & & SHAP             &          0.67 &          0.82 &          0.75 &          0.82 &          0.87 &          0.79 &          0.89 &          0.81 &          0.75 &          1.04 &          0.79 &          0.86 &          1.32 &          0.86 \\
        & & SHAP (w/ CFG)    &          0.65 &          0.80 &          0.77 &          0.75 &          0.79 &          0.77 &          0.75 &          0.76 &          0.68 &          0.78 &          0.71 &          0.80 &          0.62 &          0.74 \\
        & & IG               &          0.95 &          1.01 &          1.05 &          1.02 &          1.71 &          1.80 &          1.27 &          0.73 &          0.88 &          2.73 &          1.03 &          0.88 &          1.31 &          1.26 \\
        & & IG (w/ CFG)      &          0.94 &          1.06 &          1.19 &          0.98 &          2.31 &          5.08 &          1.29 &          0.90 &          0.86 &          4.38 &          1.14 &          1.08 &          2.34 &          1.81 \\
        & & Landmark         &          0.76 &          0.83 &          0.84 &          0.79 &          0.72 &          0.65 &          0.72 &          0.85 &          0.83 &          0.83 &          0.74 &          1.03 &          0.80 &          0.80 \\
        & & LEMON (w/o DE)   &          0.53 &          0.34 &          0.81 &          0.66 &          0.45 &          0.43 & \textbf{0.39} &          0.72 &          0.61 &          0.50 &          0.36 & \textbf{0.33} & \textbf{0.42} & \textbf{0.50} \\
        & & LEMON (w/o AP)   & \textbf{0.48} &          0.49 & \textbf{0.63} &          0.54 &          0.58 &          0.47 &          0.53 & \textbf{0.57} & \textbf{0.50} & \textbf{0.48} &          0.58 &          0.63 &          0.62 &          0.55 \\
        & & LEMON (w/o CFG)  &          0.51 & \textbf{0.29} &          0.98 &          0.70 &          0.45 & \textbf{0.42} & \textbf{0.39} &          0.81 &          0.71 & \textbf{0.48} & \textbf{0.35} & \textbf{0.33} & \textbf{0.42} &          0.53 \\
        & & LEMON            &          0.49 &          0.40 &          0.87 &          0.58 & \textbf{0.43} &          0.48 &          0.46 &          0.77 &          0.54 & \textbf{0.48} &          0.38 & \textbf{0.33} & \textbf{0.42} &          0.51 \\
        \bottomrule
    \end{tabular}
    \label{tab:perturbation-error}
\end{table*}

It is desirable that explanations are faithful to the matcher.
All useful explanations provide some simplified view of the matcher's behavior,
but we still want them to be indicative of how the matcher actually operates without being unnecessarily misleading.
Inspired by \cite{ribeiroWhyShouldTrust2016},
we make perturbations to a record pair and compare the resulting match score with what we would expect from the attributions and attribution potentials.
Specifically,
if we remove feature $i$, we expect the match score to decrease with $w_i$ (remember that $w_i$ can be negative).
Ff we inject feature $i$ into the other record, we expect the match score to increase with $p_i$.
The same applies to Landmark, but with appending instead of injecting features.
Since the baselines do not estimate attribution potentials, we ignore that perturbation for them.
We perform 1, 2, and 3 random perturbations among the interpretable features for both dual explanations.
We repeat for 500 explanations of matches and non-matches for each matcher and dataset (or all when there are less than 500 available).
Let $\delta_l$ be the set of expected match score increases and decreases for experiment $l$ out of $L$,
and let the mean absolute error be
\begin{equation}
    \mathit{MAE} = \frac{1}{L} \sum_l \Big| f(z) - \big[ f(x) + \sum_{c \in \delta_l} c \big] \Big|
\end{equation}
This error measure will favor conservative explanation methods that make small and insignificant claims,
and punish methods like Landmark and LEMON that provide higher impact explanations because of the injected/appended features.
Therefore, we define the perturbation error to be the mean absolute error by dividing by the average magnitude of the predicted change:
\begin{equation}
    \mathit{PE} = \frac{\mathit{MAE}}{\frac{1}{L} \sum_l \sum_{c \in \delta_l} |c|}
\end{equation}

Table~\ref{tab:perturbation-error} shows the perturbation error for all methods.
No method achieves truly low error levels, which is expected given the simplified assumption of independent additative attributions.
However, we observe that LEMON overall is the method with the smallest errors,
with LIME performing very similarly.
LEMON and LIME lie in the range of 0.25 to 0.75 in almost all cases,
while SHAP, IG, and Landmark often exceed 1.0.
Further, we see Landmark sometimes gets extremely high perturbation error,
especially for Magellan on the dirty datasets like Dirty iTunes-Amazon.
Upon closer inspection, we think this stems from a combination of sampling a too large neighborhood and the ineffectiveness of the double-entity generation strategy when the data does not follow the matched schemas (i.e., is dirty).

\subsection{User Study}\label{sec:user-study}

\begin{table}
    \centering
    \caption{
        Counterfactual Precision of Users After Being Shown an Explanation From LIME or LEMON.
    }
    \scriptsize
    \begin{tabular}{lcccc}
    \toprule
        \multirow{3}[3]{*}{\textbf{Dataset}} & \multicolumn{4}{c}{\textbf{Method}} \\
        \cmidrule{2-5}
        & \multicolumn{2}{c}{LIME} & \multicolumn{2}{c}{LEMON} \\
        \cmidrule(lr){2-3}
        \cmidrule(lr){4-5}
        & Match & Non-match & Match & Non-match \\
    \midrule
        \textbf{Structured} & & & & \\
        Amazon-Google      & 0.71          & 0.22 & \textbf{0.77} & \textbf{0.42} \\
        Beer               & 0.47          & 0.16 & \textbf{0.50} & \textbf{0.63} \\
        DBLP-ACM           & \textbf{0.82} & 0.06 &         0.79  & \textbf{0.23} \\
        DBLP-GoogleScholar & 0.53          & 0.08 & \textbf{0.62} & \textbf{0.27} \\
        Fodors-Zagats      & 0.59          & 0.14 & \textbf{0.69} & \textbf{0.46} \\
        iTunes-Amazon      & 0.45          & 0.18 & \textbf{0.60} & \textbf{0.60} \\
        Walmart-Amazon     & 0.55          & 0.24 & \textbf{0.63} & \textbf{0.58} \\
    \midrule
        \textbf{Dirty} & & & & \\
        DBLP-ACM           & 0.71 & 0.02 & \textbf{0.81} & \textbf{0.40} \\
        DBLP-GoogleScholar & 0.45 & 0.08 & \textbf{0.58} & \textbf{0.48} \\
        iTunes-Amazon      & 0.53 & 0.22 & \textbf{0.75} & \textbf{0.56} \\
        Walmart-Amazon     & 0.53 & 0.24 & \textbf{0.62} & \textbf{0.71} \\
    \midrule
        \textbf{Textual} & & & & \\
        Abt-Buy & 0.55 & 0.14 & \textbf{0.62} & \textbf{0.73} \\
        Company & 0.14 & 0.16 & \textbf{0.29} & \textbf{0.35} \\
    \midrule
        Mean & 0.54 & 0.15 & \textbf{0.63} & \textbf{0.49} \\
    \bottomrule
    \end{tabular}
    \label{tab:user-study}
\end{table}

To examine if explanations from LEMON improve human subjects understanding of a matcher compared to LIME,
we adopt the experiment on counterfactual interpretation from Section~\ref{sec:counterfactual-interpretation} to human subjects.
We recruit random test users from the research survey platform Prolific.
Note that these users are laymen and do not have any experience with entity matching or a background in computer science.
A user is shown an explanation for a record pair and then asked what they think would be a minimal change to the record pair that would make the matcher predict the opposite.
Each user is shown one explanation for a match and a non-match for each dataset,
and we use only the BERT-Mini matcher.
Afterward, we check what fraction of them successfully gets the opposite prediction --- i.e., the counterfactual precision.
We conduct the experiment on 50 users for LIME and 50 different users for LEMON,
and report the counterfactual precision in Table~\ref{tab:user-study}.

As expected, and in line with the experiments above,
the greatest improvement is for non-matches.
We see an average improvement in the counterfactual precision of 0.09 for matches and 0.34 for non-matches.
The results are generally less pronounced than those of the simulated experiments.
We suspect the lower maximum scores reflect the difficulty of the task for a layman,
and that the higher minimum scores reflect human ability to use common sense to make up for weak explanations.
Note that we cannot compare to Landmark~\cite{baraldiUsingLandmarksExplaining2021} since the authors do not propose any way of presenting an actual explanation to a user.
The combination of double-entity generation and not limiting $\Omega(g)$ (i.e., explaining using all features instead of limiting them to $K$) makes such a presentation non-trivial.

\subsection{Ablation Study}

Included in Table~\ref{tab:simulated-cff1} and Table~\ref{tab:perturbation-error} is also an ablation study.
We examine what happens when we remove each of the three main components of LEMON:
1) Dual Explanations. Instead of dual explanations, we produce one joint explanation for both records. We use $K = 10$ for a fair comparison.
2) Attribution Potential. We use the interpretable representation of the LIME baseline and do not estimate any attribution potential.
3) Counterfactual Granularity. We fix the granularity to be one token.
In addition, we examine the effect of adding counterfactual granularity to the baselines SHAP and integrated gradients (there is no trivial way to do the same for attribution potential).

LEMON performs better across the board for matches with dual explanations,
but the results more varied for non-matches.
This makes sense since the problematic interaction effects mainly occur when two records match and have a lot of similar content.
Unsurprisingly, since its primary goal is to explain how records could match better, attribution potential only significantly improves non-match explanations.
Nevertheless, the improvement for non-matches is dramatic,
demonstrating how effective attribution potential is for explaining record pairs that do not match.
Finally, we observe that the effectiveness of counterfactual granularity varies greatly from dataset to dataset.
It makes the most difference on datasets where we consider the records to have multiple high-quality pieces of information --- either in the form of several high-quality attributes or long textual attributes with multiple high-quality keywords.

\subsection{Stability}\label{sec:stability}
\begin{figure*}
    \centering
    \includegraphics{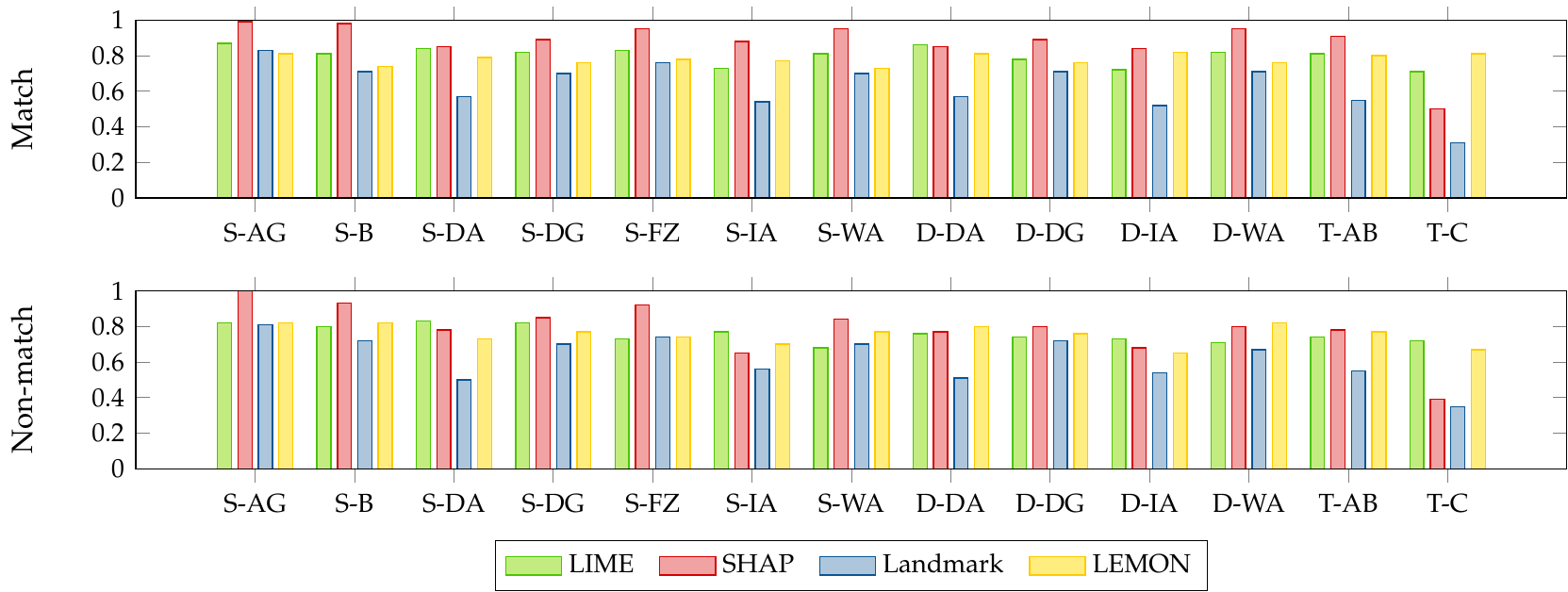}
    \caption{
        Stability of explainability methods based on neighborhood sampling for BERT-Mini on all datasets.
    }
    \label{fig:stability}
\end{figure*}

Several of the benchmarked methods, including LEMON itself, rely on random sampling of the neighborhood of $x$.
Different initial random seeds will result in different explanations.
However, with sufficient samples we would like a well-behaved method to generate similar explanations --- i.e. explanations to be stable and not change much if different random seeds are used.
Stability is a desirable trait from a trust perspective,
but also especially useful when examining or debugging a matcher.
If we make changes to a matcher, we want to be confident that the differences we observe in the explanations mostly reflect the matcher changes and not instability of the explanation method.
LEMON not only relies on sampling the neighborhood of $x$ in the interpretable domain $I_x$,
but also on sampling to approximate $t_x$ when translating from $I_x$.
A natural question to ask is if this additional random sampling hurts stability.

Let $e_x^1 = \{(w_{i1},p_{i2})\}_{i \in E_1}$ and $e_x^2 = \{(w_{i2},p_{i2})\}_{i \in E_2}$ be two explanations for the same input $x$ and matcher $f$ with a different random seed.
Let the the similarity between the two explanations $s(e_1, e_2) = \frac{e_1 \cap e_2}{e_1 \cup e_2}$ be the weighted Jaccard coefficient such that the intersection is
\begin{equation}
        e_x^1 \cap e_x^2 = \sum_{i \in E_1 \cap E_2} \Big[ (w_{i1} \dot{\cap} w_{i2}) + (p_{i1} \dot{\cap} p_{i2}) \Big]
\end{equation}
and the union is
\begin{equation}
\begin{split}
    e_x^1 \cup e_x^2 =& \sum_{i \in E_1 \cap E_2} \max(|w_{i1}|, |w_{i2}|) + \max(|p_{i1}|, |p_{i2}|) \\
    & + \sum_{i \in E_1 \setminus E_2} (w_{i1} + p_{i1})
    + \sum_{i \in E_2 \setminus E_1} (w_{i2} + p_{i2})
\end{split}
\end{equation}
where $\dot{\cap}$ is a shorthand for
\begin{equation}
    r \dot{\cap} q = H(r q) \min(|r|, |q|)
\end{equation}
and $H$ is the unit step function.
For LIME and SHAP, $p_i$ is $0$ for all $i$.
To be able to compare explanations of different granularity, all explanations are normalized to single token interpretable features --- i.e. if feature $i$ is an $n$-token interpretable feature we split it into $n$ features with attribution $\frac{w_i}{n}$ and attribution potential $\frac{p_i}{n}$.
Finally,
we define the stability of an explanation method as the expected similarity between two explanations $\mathbb{E}_{x} \big[ s(e_x^1, e_x^2) \big]$.
Note that this definition slightly favors methods such as SHAP and Landmark that uses all interpretable features in its explanations instead of only the $K$ most important like LIME and LEMON.
Picking the $K$ most important interpretable features controls the explanation complexity at the cost of exposing the method to more instability because small changes in importance can change which features are within or outside top $K$.

Figure~\ref{fig:stability} shows the estimated stability of LIME, SHAP, Landmark, and LEMON for BERT-Mini on all datasets.
For each dataset, we sample 100 predicted matches and non-matches uniformly at random, generate two explanations with different random seed for each example, and average the similarities.
We see that LEMON is relatively stable and is similar to LIME in terms of stability.
This is important because it shows LEMON does not degrade in stability despite the sampling-based approximation of $t_x$.
SHAP is overall the most stable method,
while Landmark is the least stable.
To understand these differences in stability it is important to also take into account the sample size.

\subsection{Neighborhood Sample Size}\label{sec:neighborhood-sample-size}
\begin{figure}
    \centering
    \includegraphics{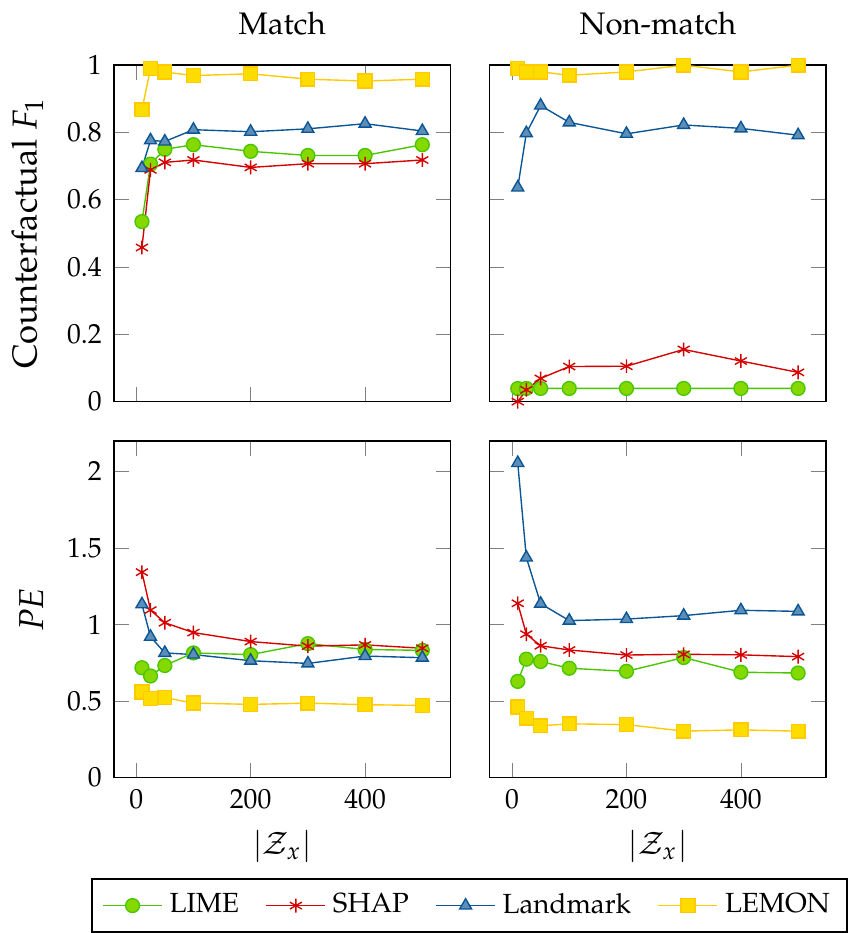}
    \caption{
        Counterfactual $F_1$ and perturbation error ($\mathit{PE}$) of explainability methods based on neighborhood sampling for BERT-Mini on Abt-Buy when varying the neighborhood sampling size $|\mathcal{Z}_x|$.
    }
    \label{fig:vary-num-samples}
\end{figure}
\begin{figure}
    \centering
    \includegraphics{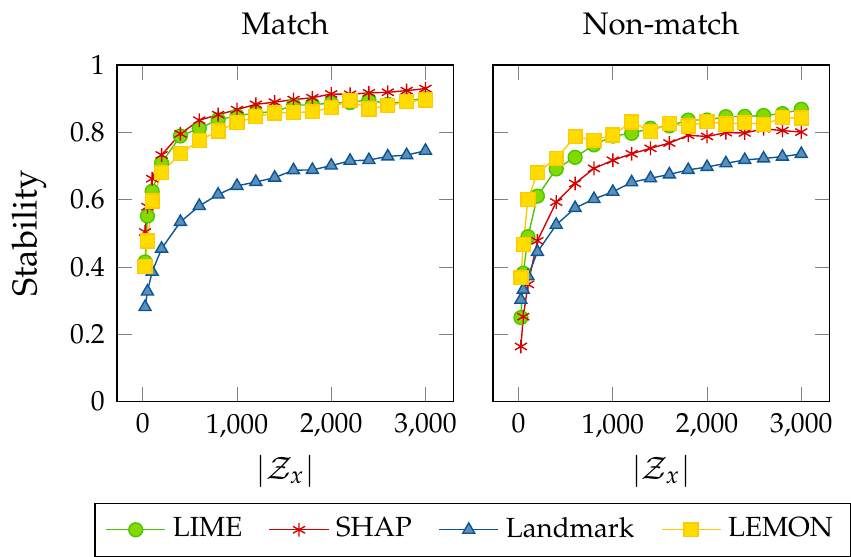}
    \caption{
        Stability of explainability methods based on neighborhood sampling for BERT-Mini on the Abt-Buy dataset when varying the neighborhood sampling size $|\mathcal{Z}_x|$.
    }
    \label{fig:stability-vary-num-samples}
\end{figure}

From our experience, the main concern when choosing the neighborhood sample size $|\mathcal{Z}_x|$ is stability.
From Figure~\ref{fig:vary-num-samples} we see that one achieves satisfactory counterfactual $F_1$ score and perturbation error with relatively few samples,
but as we will see, it takes considerably more samples to get stable explanations.
Thus, picking $|\mathcal{Z}_x|$ mostly boils down to a trade-off between stability and speed (see \ref{sec:runtime} for a discussion about runtime).

The different neighborhood sampling-based methods have different strategies for picking a sample size\footnote{SHAP defaults to $2 d_x + 2048$, Landmark to $500$, and our LIME baseline and LEMON to $\max(500, \min(30 d_x, 3000))$.},
so it is interesting to compare the stability at equal sample sizes.
Figure~\ref{fig:stability-vary-num-samples} shows the stability of the explainability methods for BERT-Mini on the Abt-Buy dataset when we vary the neighborhood sample size.
As before, we sample 100 predicted matches and non-matches, generate two explanations per example, and estimate the stability to be the average similarity between the explanation pairs.
We observe that LEMON is close to or equally sample efficient as LIME and SHAP for matches,
and slightly more for non-matches.
This shows that the difference in stability between SHAP and LEMON is mainly a matter of difference in sample sizes.
We deliberately use a less aggressive sampling scheme for LEMON than SHAP because we find the returns in terms of stability diminishing --- especially given the higher computational footprint of LEMON.
Landmark's instability, however, can not be attributed to the lower sampling size.
It is clear from Figure~\ref{fig:stability-vary-num-samples} that the method is significantly less sampling efficient than the others.
We suspect this is mostly due to the large neighborhood used when sampling.

\subsection{Explanation Complexity}\label{sec:explanation-complexity}
\begin{figure}
    \centering
    \includegraphics{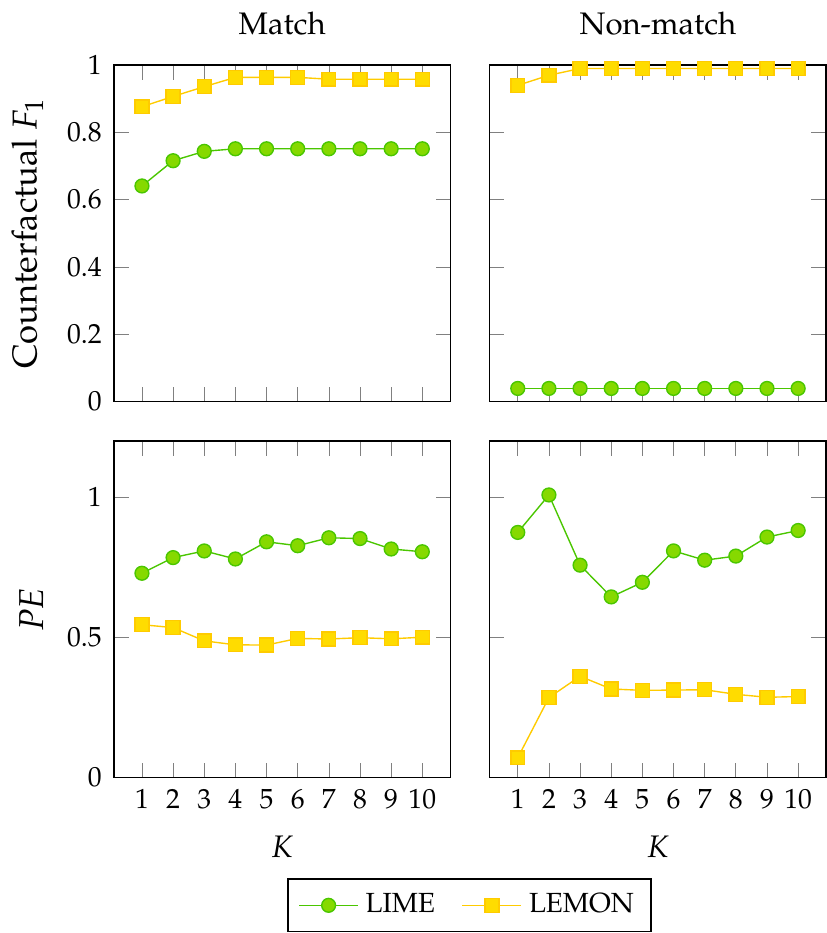}
    \caption{
        Counterfactual $F_1$ and perturbation error ($\mathit{PE}$) of explainability methods based on neighborhood sampling for BERT-Mini on Abt-Buy when varying $K$.
    }
    \label{fig:vary-k}
\end{figure}

An important distinction between LEMON and Landmark is that LEMON, as LIME, limits the explanation complexity $\Omega(g)$ by constraining the number of interpretable features used in an explanation to $K < d_x$.
This is important because we can not generally expect users to consume explanations with a large number of interpretable components.
We consider $K = 5$ default for LEMON and have used this for all experiments,
since we consider this a reasonable number of features for user consumption in practice.
Furthermore, we argue that the choice of $K$ is indeed mostly a matter of what is practical to the user.
Figure~\ref{fig:vary-k} shows how the counterfactual $F_1$ score and perturbation error vary depending on the choice of $K$ for LIME and LEMON (remember SHAP and Landmark use all interpretable features).
We see that for $K \geq 3$ the counterfactual interpretation and explanation faithfulness is not affected much by the choice of $K$.
For very low values of $K$ we lose the necessary expressive power needed to capture the matcher's behavior --- which makes it hard to produce counterfactually interpretable explanations.

\subsection{Runtime}\label{sec:runtime}
\begin{figure}
    \centering
    \includegraphics{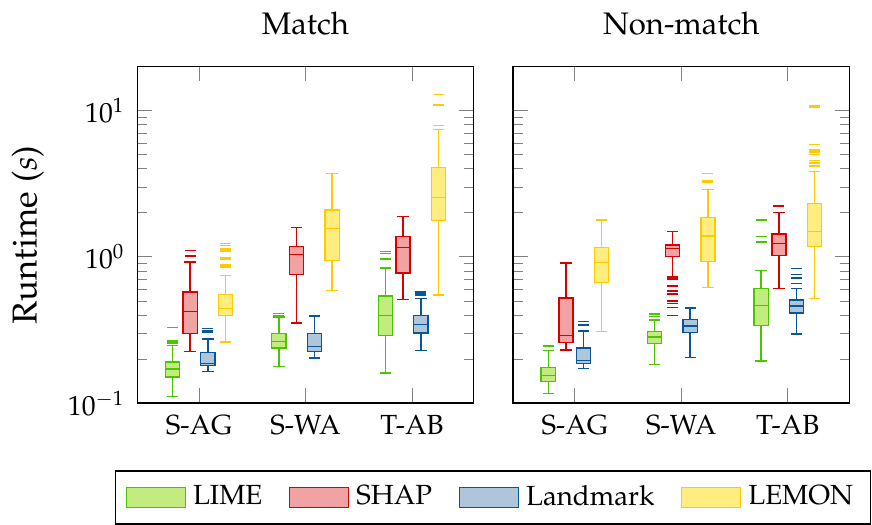}
    \caption{
        Runtime of explainability methods based on neighborhood sampling for BERT-Mini on three different datasets.
    }
    \label{fig:runtime}
\end{figure}
\begin{figure}
    \centering
    \includegraphics{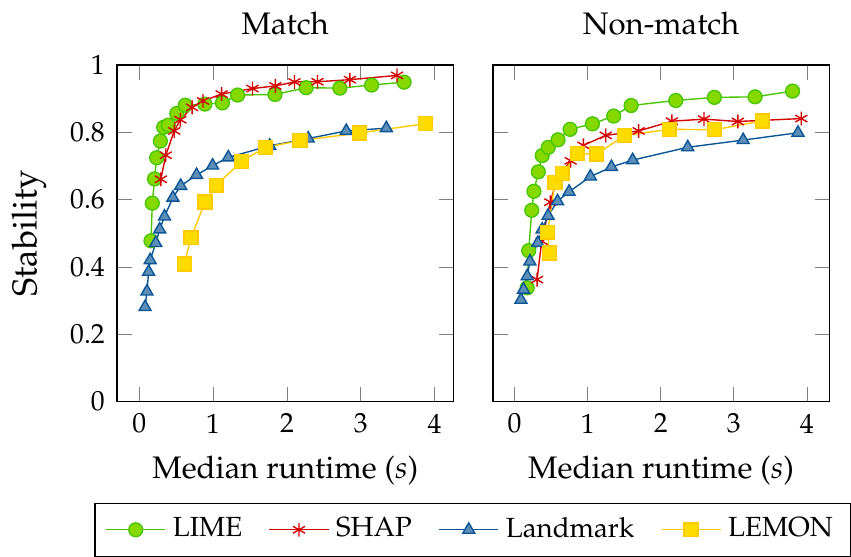}
    \caption{
        Stability of explainability methods based on neighborhood sampling for BERT-Mini on Abt-Buy when varying the median runtime (by changing $|\mathcal{Z}_x|$).
    }
    \label{fig:runtime-stability}
\end{figure}

One of the main disadvantages of local post hoc neighborhood sampling methods are long runtimes.
This is a result of having to do inference on a large number of sampled inputs.
Of course, LEMON is more prone to this than existing work due to the approximation of $t_x$ and counterfactual granularity.
Figure~\ref{fig:runtime} shows the time needed to make a single explanation of a BERT-Mini matcher prediction for three different datasets on a NVIDIA RTX 2080 Ti for the different neighborhood sampling-based methods.
Each boxplot shows the distribution of 100 explanations.
LEMON generally takes the longest time, with SHAP being most comparable.
Note that the runtime varies significantly for every methods even on the same dataset.
This is because inference time depends heavily on the input size, which varies between record pairs and depend on the random perturbation.
LEMON's runtime varies more because of how the counterfactual granularity is found.

While the runtime is longer than in previous work, we argue it is still within reason for most applications on most datasets --- especially taking into consideration the improvement in explanation quality seen in Section~\ref{sec:counterfactual-interpretation}, \ref{sec:explanation-faithfulness}, and \ref{sec:user-study}.
Moreover, it is possible to trade off some stability for shorter runtime if desired.
As mentioned in Section~\ref{sec:neighborhood-sample-size}, the choice of $|\mathcal{Z}_x|$ is essentially a trade-off between stability and speed.
Figure~\ref{fig:runtime-stability} plots the stability against the median runtime for explaining BERT-Mini on the Abt-Buy dataset (one of the datasets with the longest runtime) when we vary $|\mathcal{Z}_x|$.
We see that LEMON has a stability-runtime trade-off comparable to Landmark.
By reducing the neighborhood sample size we can achieve more similar runtime to Landmark at the expense of also getting similar (low) stability as Landmark.
To what degree depends on the dataset, but there is significant flexibility if lower runtime is critical.

\vspace{-0.5em}
\section{Conclusion}

Local post hoc feature attribution is a valuable and popular type of explainability method that can explain any classifier,
but standard methods leave significant room for improvement when applied to entity matching.
We have identified three challenges of applying such methods to entity matching and proposed LEMON,
a model-agnostic and schema-flexible method that addresses all three challenges.
Experiments and a novel evaluation method for explainable entity matching show that our proposed method is more faithful to the matcher and more effective in explaining to the user where the decision boundary is --- especially for non-matches.
Lastly, user studies support a real-world improvement in understanding for a layman seeing LEMON explanations compared to naive LIME explanations.

There is still much to be done within explainable entity matching.
A disadvantage of LEMON (and other perturbation-based methods like LIME, SHAP, and Landmark) is their running time.
Even though it is possible to trade off significantly shorter running time for explanation stability,
and trivial to parallelize the computational bottleneck (running inference of matcher $f$),
depending on the hardware and matcher,
it might still be infeasible in practice to do real-time explanation or generate explanations for all record pairs in large datasets,
while still achieving satisfactory stability.
Therefore, more efficient sampling strategies should be explored.
Furthermore, there is more to be done on examining the adaptation of other explainability methods for entity matching in-depth,
and on how to evaluate them.
Our experiments and ablation study show that dual explanations and counterfactual granularity are easily applicable to SHAP and gradient-based methods,
and that they are indeed effective for other methods than LIME.
It is less clear how one would adapt the ideas of attribution potential to those methods,
and we hope to address that in the future.

\ifCLASSOPTIONcompsoc
  \section*{Acknowledgments}
\else
  \section*{Acknowledgment}
\fi

This work is supported by Cognite and the Research Council of Norway under Project 298998.
We thank the reviewers, Hassan Abedi Firouzjaei, Yanzhe Bekkemoen, Jon Atle Gulla, Dhruv Gupta, Benjamin Kille, Ludvig Killingberg, Kjetil Nørvåg, Mateja Stojanović, and Bjørnar Vassøy for valuable feedback.

\ifCLASSOPTIONcaptionsoff
  \newpage
\fi



%



\bibliographystyle{IEEEtran}
\vspace{-1em}
\bibliography{references}

%

\vspace{-2.5em}
\begin{IEEEbiography}[
{\includegraphics[width=1in,height=1.25in,clip,keepaspectratio]{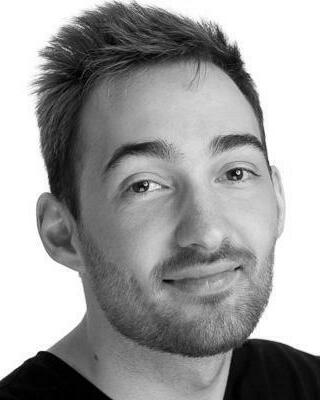}}
]{Nils Barlaug}
is working towards the PhD degree at the Norwegian University of Science and Technology in cooperation with Cognite.
His research interests include data integration, machine learning, and explainability.
\end{IEEEbiography}


\vfill


\clearpage

\appendices
\section{Other Matchers}\label{sec:other-matchers}
\FloatBarrier
In addition to Magellan and BERT-Mini,
it is also interesting to evaluate LEMON on larger transformer models and other deep learning architectures.
To that end we perform the experiments on counterfactual interpretation and explanation faithfulness from Section~\ref{sec:counterfactual-interpretation} and \ref{sec:explanation-faithfulness} on a RoBERTa-based~\cite{liuRoBERTaRobustlyOptimized2019} baseline DITTO model and DeepMatcher~\cite{mudgalDeepLearningEntity2018}.

\subsection{DeepMatcher}
The authors \cite{mudgalDeepLearningEntity2018} explore a range of different deep learning models for entity matching.
We use their hybrid model since it performs the best overall.
Each model is trained for 15 epochs with a batch size of 32 and a negative to positive sampling ratio of 3.
The model is evaluated on the validation set after every epoch and the best model is kept.
Note that we do not perform an exhaustive hyperparameter search like the authors and instead use default settings as provided by the publicly available implementation\footnote{\url{https://github.com/anhaidgroup/deepmatcher}} from the authors --- which gives performance reasonably close to what they report.

\subsection{RoBERTa}
The authors of DITTO~\cite{liDeepEntityMatching2020} evaluated a number of prominent transformer-based natural language models and found RoBERTa~\cite{liuRoBERTaRobustlyOptimized2019} to generally perform the best.
We train each baseline RoBERTa-based DITTO model the same way as described in Section~\ref{sec:matchers} for BERT-Mini.

\subsection{Results}
Table~\ref{tab:models} shows the performance of the DeepMatcher and RoBERTa models on every dataset.
We have also repeated the performance of Magellan and BERT-Mini for easy comparison.
As we see,
DeepMatcher generally outperforms Magellan on the dirty and textual datasets while the results are more mixed on the structured datasets --- which is in line with the DeepMatcher authors' reported results \cite{mudgalDeepLearningEntity2018}.
Furthermore, BERT-Mini performs better than DeepMatcher on most datasets while RoBERTa performs even better than BERT-Mini.
This shows that even though bigger transformer models are better,
a conservatively sized model is able to outperform the previous generation deep learning method.

Figure~\ref{tab:simulated-cff1-roberta-dm} shows the counterfactual $F_1$ score for LIME, SHAP, Landmark, and LEMON for DeepMatcher and RoBERTa across all datasets.
We see that the results are similar to those of Magellan and BERT-Mini in Section~\ref{sec:counterfactual-interpretation},
and the biggest improvements over the baselines are seen for non-matches.
The results further strengthen the claim that LEMON is model-agnostic by showing that it is equally functional for other deep learning architectures and even bigger transformer models.

Furthermore, Figure~\ref{tab:simulated-pe-roberta-dm} shows the perturbation error $\mathit{PE}$ for the same explainability methods and matchers across all datasets.
The general tendencies are the same as for Magellan and BERT-Mini in Section~\ref{sec:explanation-faithfulness}.
LEMON is overall similar to LIME (but performs noticeably worse on some datasets for non-matches with DeepMatcher), while still being significantly more faithful than SHAP and Landmark.
We note that even though Landmark has substantially higher perturbation error than LEMON and LIME for DeepMatcher and RoBERTa,
it is still considerably better than for Magellan.
We are uncertain why Magellan triggers particularly large errors,
but we suspect it is because Magellan has a less forgiving decision boundary that changes more abruptly when multiple attributes are perturbed at the same time since it uses per-attribute string similarity metrics.

\begin{table}
    \centering
    \caption{
    $F_1$ Score for the Magellan, DeepMatcher, BERT-Mini, and RoBERTa Matchers Used in the Experiments on the Public DeepMatcher~\cite{mudgalDeepLearningEntity2018} Benchmark Dataset.
    }
    \scriptsize
    \begin{tabular}{llcccc}
    \toprule
        \multirow{2}[1]{*}{\textbf{Type}} & \multirow{2}[1]{*}{\textbf{Name}} & \multicolumn{4}{c}{\textbf{Matcher $F_1$}} \\
        \cmidrule{3-6}
        & & MG & DM & BM & RoBERTa \\
    \midrule
        \multirow{7}{*}{Structured}
        & Amazon-Google  & 0.52 & 0.67 & 0.67 & 0.72 \\
        & Beer           & 0.85 & 0.69 & 0.76 & 0.90 \\
        & DBLP-ACM       & 0.99 & 0.98 & 0.98 & 0.99 \\
        & DBLP-Scholar   & 0.94 & 0.95 & 0.93 & 0.95 \\
        & Fodors-Zagats  & 1.00 & 0.91 & 0.95 & 1.00 \\
        & iTunes-Amazon  & 0.90 & 0.87 & 0.93 & 0.94 \\
        & Walmart-Amazon & 0.66 & 0.66 & 0.80 & 0.87 \\
        \midrule
        \multirow{4}{*}{Dirty}
        & DBLP-ACM       & 0.91 & 0.96 & 0.97 & 0.99 \\
        & DBLP-Scholar   & 0.83 & 0.92 & 0.94 & 0.95 \\
        & iTunes-Amazon  & 0.53 & 0.65 & 0.90 & 0.96 \\
        & Walmart-Amazon & 0.41 & 0.39 & 0.79 & 0.86 \\
        \midrule
        \multirow{2}{*}{Textual}
        & Abt-Buy & 0.51 & 0.68 & 0.81 & 0.88 \\
        & Company & 0.57 & 0.89 & 0.91 & 0.91 \\
    \bottomrule
    \end{tabular}
    \label{tab:models}
\end{table}

\begin{table*}
    \centering
    \caption{
        Counterfactual $F_1$ Score for All the Evaluated Explainability Methods Across All Datasets and the Two Matchers.
    }
    \scriptsize
    \begin{tabular}{cclcccccccccccccc}
        \toprule
        \multirow{3}{*}[-0.4em]{\textbf{Model}} & \multirow{3}{*}[-0.4em]{\textbf{Type}} & \multirow{3}{*}[-0.4em]{\textbf{Method}} & \multicolumn{13}{c}{\textbf{Dataset}} & \\
        & & & \multicolumn{7}{c}{Structured} & \multicolumn{4}{c}{Dirty} & \multicolumn{2}{c}{Textual} \\
        \cmidrule(lr){4-10} \cmidrule(lr){11-14} \cmidrule(lr){15-16}
        & & & AG & B & DA & DG & FZ & IA & WA & DA & DG & IA & WA & AB & C & Mean \\
        \midrule
         \multirow{8}{*}{DeepMatcher} & \multirow{4}{*}{Match}
          & LIME             &         0.98 &          0.76 &          0.86 &          0.68 &          0.74 &          0.54 &          0.84 &          0.77 &          0.85 &          0.59 &          0.91 &          0.93 &          0.22 &          0.74 \\
        & & SHAP             &         0.96 &          0.76 & \textbf{1.00} &          0.81 &          0.98 & \textbf{0.60} &          0.86 &          0.99 &          0.97 & \textbf{0.62} &          0.92 &          0.96 & \textbf{0.35} &          0.83 \\
        & & Landmark         &         0.98 &          0.76 &          0.95 &          0.90 &          0.96 &          0.57 &          0.92 &          0.96 &          0.95 &          0.52 &          0.85 &          0.90 &          0.12 &          0.79 \\
        & & LEMON            &\textbf{1.00} & \textbf{0.80} & \textbf{1.00} & \textbf{1.00} & \textbf{1.00} &          0.57 & \textbf{0.98} & \textbf{1.00} & \textbf{1.00} &          0.54 & \textbf{1.00} & \textbf{0.99} &          0.21 & \textbf{0.85} \\
        \cmidrule(l){2-17}
        & \multirow{4}{*}{Non-match}
          & LIME             &         0.18 &          0.26 &          0.02 &          0.03 &          0.00 &          0.22 &          0.10 &          0.02 &          0.07 &          0.11 &          0.14 &          0.54 &          0.03 &          0.13 \\
        & & SHAP             &         0.22 &          0.30 &          0.00 &          0.01 &          0.01 &          0.60 &          0.17 &          0.00 &          0.02 &          0.35 &          0.13 &          0.51 &          0.08 &          0.18 \\
        & & Landmark         &         0.25 &          0.74 & \textbf{0.17} &          0.14 &          0.64 &          0.37 &          0.15 &          0.10 &          0.21 &          0.54 &          0.57 & \textbf{0.98} &          0.04 &          0.38 \\
        & & LEMON            &\textbf{0.62} & \textbf{0.97} &          0.15 & \textbf{0.76} & \textbf{0.89} & \textbf{0.69} & \textbf{0.89} & \textbf{0.73} & \textbf{0.76} & \textbf{0.99} & \textbf{0.65} & \textbf{0.98} & \textbf{0.90} & \textbf{0.77} \\
        \midrule
         \multirow{8}{*}{RoBERTa} & \multirow{4}{*}{Match}
          & LIME             &         0.99 &          0.80 & \textbf{1.00} &          0.89 &          0.67 &          0.87 &          0.77 &          0.99 &          0.77 &          0.84 &          0.71 &          0.79 &          0.19 &          0.79 \\
        & & SHAP             &         0.99 &          0.83 & \textbf{1.00} &          0.99 &          0.73 & \textbf{1.00} &          0.63 & \textbf{1.00} &          0.98 &          0.87 &          0.68 &          0.81 & \textbf{0.27} &          0.83 \\
        & & Landmark         &         0.99 & \textbf{1.00} & \textbf{1.00} &          0.88 &          0.82 &          0.92 &          0.82 &          0.99 &          0.80 &          0.86 &          0.82 &          0.94 &          0.06 &          0.84 \\
        & & LEMON            &\textbf{1.00} & \textbf{1.00} & \textbf{1.00} & \textbf{1.00} & \textbf{1.00} & \textbf{1.00} & \textbf{0.95} & \textbf{1.00} & \textbf{1.00} & \textbf{0.96} & \textbf{0.96} & \textbf{0.99} & \textbf{0.27} & \textbf{0.93} \\
        \cmidrule(l){2-17}
        & \multirow{4}{*}{Non-match}
          & LIME             &         0.10 &          0.15 &          0.00 &          0.05 &          0.00 &          0.05 &          0.04 &          0.00 &          0.02 &          0.09 &          0.10 &          0.04 &          0.04 &          0.05 \\
        & & SHAP             &         0.04 &          0.16 &          0.00 &          0.02 &          0.00 &          0.00 &          0.12 &          0.00 &          0.00 &          0.05 &          0.16 &          0.04 &          0.06 &          0.05 \\
        & & Landmark         &         0.23 &          0.18 &          0.09 &          0.13 &          0.01 &          0.11 &          0.35 &          0.05 &          0.11 &          0.13 &          0.54 &          0.58 &          0.01 &          0.19 \\
        & & LEMON            &\textbf{0.73} & \textbf{0.58} & \textbf{0.18} & \textbf{0.71} & \textbf{0.83} & \textbf{0.87} & \textbf{0.79} & \textbf{0.53} & \textbf{0.77} & \textbf{0.75} & \textbf{0.92} & \textbf{0.97} & \textbf{0.94} & \textbf{0.74} \\
        \bottomrule
    \end{tabular}
    \label{tab:simulated-cff1-roberta-dm}
\end{table*}

\begin{table*}
    \centering
    \caption{
        Perturbation Error $\mathit{PE}$ for All the Evaluated Explainability Methods Across All Datasets and the Two Matchers.
    }
    \scriptsize
    \begin{tabular}{cclcccccccccccccc}
        \toprule
        \multirow{3}{*}[-0.4em]{\textbf{Model}} & \multirow{3}{*}[-0.4em]{\textbf{Type}} & \multirow{3}{*}[-0.4em]{\textbf{Method}} & \multicolumn{13}{c}{\textbf{Dataset}} & \\
        & & & \multicolumn{7}{c}{Structured} & \multicolumn{4}{c}{Dirty} & \multicolumn{2}{c}{Textual} \\
        \cmidrule(lr){4-10} \cmidrule(lr){11-14} \cmidrule(lr){15-16}
        & & & AG & B & DA & DG & FZ & IA & WA & DA & DG & IA & WA & AB & C & Mean \\
        \midrule
        \multirow{8}{*}{DeepMatcher} & \multirow{4}{*}{Match}
          & LIME             &\textbf{0.36} & \textbf{0.24} & \textbf{0.34} &          0.46 &          0.29 & \textbf{0.17} &          0.55 &          0.55 &          0.46 & \textbf{0.23} &          0.42 & \textbf{0.30} &          1.28 & \textbf{0.43} \\
        & & SHAP             &         0.70 &          0.34 &          0.82 &          0.96 &          0.41 &          0.40 &          0.94 &          0.93 &          1.09 &          0.34 &          0.70 &          0.46 &          1.35 &          0.73 \\
        & & Landmark         &         0.82 &          1.38 &          0.65 &          0.91 &          0.38 &          0.68 &          0.84 &          1.01 &          1.13 &          2.95 &          1.75 &          1.17 &          1.11 &          1.14 \\
        & & LEMON            &         0.38 &          0.69 &          0.35 & \textbf{0.44} & \textbf{0.23} &          0.28 & \textbf{0.37} & \textbf{0.46} & \textbf{0.43} &          0.57 & \textbf{0.38} &          0.41 & \textbf{0.62} & \textbf{0.43} \\
        \cmidrule(l){2-17}
        & \multirow{4}{*}{Non-match}
          & LIME             &\textbf{0.48} & \textbf{0.36} & \textbf{0.32} & \textbf{0.41} & \textbf{0.22} & \textbf{0.22} & \textbf{0.52} & \textbf{0.44} &          0.50 & \textbf{0.24} & \textbf{0.49} & \textbf{0.39} &          0.70 & \textbf{0.41} \\
        & & SHAP             &         0.75 &          0.65 &          0.83 &          0.73 &          0.44 &          0.49 &          0.79 &          0.92 &          0.89 &          0.57 &          0.83 &          0.75 &          0.92 &          0.74 \\
        & & Landmark         &         0.80 &          1.02 &          0.85 &          0.82 &          0.44 &          0.63 &          0.82 &          0.83 &          0.80 &          1.42 &          0.90 &          2.21 &          1.28 &          0.99 \\
        & & LEMON            &         0.56 &          0.91 &          0.69 &          0.55 &          0.30 &          0.51 &          0.53 &          0.49 & \textbf{0.48} &          0.78 &          0.58 &          0.47 & \textbf{0.45} &          0.56 \\
        \midrule
         \multirow{8}{*}{RoBERTa} & \multirow{4}{*}{Match}
          & LIME             &\textbf{0.34} & \textbf{0.33} & \textbf{0.39} &          0.51 &          0.60 & \textbf{0.41} &          0.67 & \textbf{0.36} &          0.54 &          0.54 &          0.72 &          0.93 &          1.34 &          0.59 \\
        & & SHAP             &         0.88 &          1.02 &          1.27 &          1.15 &          0.91 &          1.05 &          1.32 &          1.12 &          1.07 &          1.27 &          1.37 &          0.94 &          1.62 &          1.15 \\
        & & Landmark         &         0.82 &          0.73 &          0.79 &          0.88 &          0.84 &          0.78 &          0.86 &          0.80 &          0.89 &          0.93 &          0.90 &          0.68 &          1.33 &          0.86 \\
        & & LEMON            &         0.36 &          0.43 & \textbf{0.39} & \textbf{0.47} & \textbf{0.56} &          0.44 & \textbf{0.53} & \textbf{0.36} & \textbf{0.50} & \textbf{0.53} & \textbf{0.52} & \textbf{0.50} & \textbf{0.54} & \textbf{0.47} \\
        \cmidrule(l){2-17}
        & \multirow{4}{*}{Non-match}
          & LIME             &         0.55 & \textbf{0.64} & \textbf{0.40} & \textbf{0.47} &          0.57 & \textbf{0.39} &          0.91 &          0.92 &          0.61 &          0.74 &          0.79 &          0.79 &          0.49 &          0.64 \\
        & & SHAP             &         0.88 &          1.05 &          1.09 &          1.31 & \textbf{0.49} &          0.89 &          0.96 & \textbf{0.63} &          0.89 &          1.71 &          0.96 &          0.82 &          1.03 &          0.98 \\
        & & Landmark         &         0.81 &          0.85 &          0.81 &          0.80 &          0.83 &          0.79 &          0.78 &          0.81 &          0.79 &          0.77 &          0.78 &          0.81 &          1.00 &          0.82 \\
        & & LEMON            &\textbf{0.53} &          0.69 &          1.05 &          0.56 &          0.74 &          0.61 & \textbf{0.56} &          0.74 & \textbf{0.58} & \textbf{0.57} & \textbf{0.44} & \textbf{0.31} & \textbf{0.42} & \textbf{0.60} \\
        \bottomrule
    \end{tabular}
    \label{tab:simulated-pe-roberta-dm}
\end{table*}

\section{Extensive Results}\label{sec:extensive-results}
\subsection{Precision-Recall Trade-off}

Table~\ref{tab:simulated-cff1} from Section~\ref{sec:counterfactual-interpretation} reports counterfactual $F_1$ scores.
For completeness we also present the counterfactual precision and recall for those same experiments in Table~\ref{tab:simulated-cff1-precision-recall}.
The desired trade-off between precision and recall will depend on the use case,
so we acknowledge that $F_1$ score will never be a perfect metric.
One could argue that counterfactual precision is often more important than counterfactual recall because it is harder to trust explanations that convey false information than explanations that fail to convey anything useful.
However, it would still be challenging to define exactly what the trade-off should be.
Regardless,
we see that all evaluated methods have relatively high precision and in general higher precision than recall.

As expected,
we see that the main reason the baselines perform badly on non-matches is that the counterfactual recall is low.
In other words,
they simply struggle to generate explanations that could be interpreted counterfactually.

\begin{table*}[h]
    \centering
    \caption{
        Counterfactual Precision and Recall for the Reported $F_1$ Scores in Table~\ref{tab:simulated-cff1}.
    }
    \tiny
    \begin{tabular}{
        c
        c
        l
        c @{\hskip 3pt}
        c @{\hskip 6pt}
        c @{\hskip 3pt}
        c @{\hskip 6pt}
        c @{\hskip 3pt}
        c @{\hskip 6pt}
        c @{\hskip 3pt}
        c @{\hskip 6pt}
        c @{\hskip 3pt}
        c @{\hskip 6pt}
        c @{\hskip 3pt}
        c @{\hskip 6pt}
        c @{\hskip 3pt}
        c @{\hskip 6pt}
        c @{\hskip 3pt}
        c @{\hskip 6pt}
        c @{\hskip 3pt}
        c @{\hskip 6pt}
        c @{\hskip 3pt}
        c @{\hskip 6pt}
        c @{\hskip 3pt}
        c @{\hskip 6pt}
        c @{\hskip 3pt}
        c @{\hskip 6pt}
        c @{\hskip 3pt}
        c @{\hskip 6pt}
        c @{\hskip 3pt}
        c @{\hskip 6pt}
    }
        \toprule
        \multirow{3}{*}[-0.4em]{\textbf{Model}} & \multirow{3}{*}[-0.4em]{\textbf{Type}} & \multirow{3}{*}[-0.4em]{\textbf{Method}} & \multicolumn{26}{@{}c}{\textbf{Dataset}} & \\
        & & & \multicolumn{14}{@{}c}{Structured} & \multicolumn{8}{@{}c}{Dirty} & \multicolumn{4}{@{}c}{Textual} \\
        \cmidrule(lr{6pt}){4-17} \cmidrule(l{0pt}r{6pt}){18-25} \cmidrule(l{0pt}r{6pt}){26-29}
        & & & \multicolumn{2}{c}{AG} & \multicolumn{2}{@{}c}{B} & \multicolumn{2}{@{}c}{DA} & \multicolumn{2}{@{}c}{DG} & \multicolumn{2}{@{}c}{FZ} & \multicolumn{2}{@{}c}{IA} & \multicolumn{2}{@{}c}{WA} & \multicolumn{2}{@{}c}{DA} & \multicolumn{2}{@{}c}{DG} & \multicolumn{2}{@{}c}{IA} & \multicolumn{2}{@{}c}{WA} & \multicolumn{2}{@{}c}{AB} & \multicolumn{2}{@{}c}{C} & \multicolumn{2}{@{}c}{Mean} \\
        & & & P & R & P & R & P & R & P & R & P & R & P & R & P & R & P & R & P & R & P & R & P & R & P & R & P & R & P & R \\
        \midrule
         \multirow{16}{*}{Magellan} & \multirow{7}{*}{Match}
          & LIME             &0.96 & 0.96 & 0.88 & 0.79 & 1.00 & 1.00 & 0.96 & 0.80 & 0.77 & 0.77 & 1.00 & 0.97 & 1.00 & 0.70 & 0.72 & 0.38 & 0.88 & 0.72 & 1.00 & 0.82 & 0.94 & 0.80 & 0.96 & 0.95 & 0.67 & 0.36 & 0.90 & 0.77 \\
        & & SHAP             &0.95 & 0.95 & 1.00 & 1.00 & 1.00 & 1.00 & 1.00 & 1.00 & 0.95 & 0.95 & 1.00 & 1.00 & 0.99 & 0.71 & 0.65 & 0.65 & 0.96 & 0.96 & 0.68 & 0.68 & 0.89 & 0.75 & 0.98 & 0.98 & 0.79 & 0.60 & 0.91 & 0.86 \\
        & & SHAP (w/ CFG)    &1.00 & 1.00 & 1.00 & 1.00 & 1.00 & 1.00 & 1.00 & 1.00 & 1.00 & 1.00 & 1.00 & 1.00 & 0.98 & 0.98 & 0.99 & 0.99 & 1.00 & 1.00 & 0.91 & 0.91 & 0.99 & 0.99 & 0.99 & 0.99 & 0.98 & 0.78 & 0.99 & 0.97 \\
        & & Landmark         &0.92 & 0.92 & 0.89 & 0.89 & 1.00 & 1.00 & 0.96 & 0.96 & 0.73 & 0.73 & 0.87 & 0.87 & 0.99 & 0.91 & 1.00 & 0.60 & 0.80 & 0.68 & 0.90 & 0.86 & 0.90 & 0.89 & 0.95 & 0.95 & 0.29 & 0.27 & 0.86 & 0.81 \\
        & & LEMON (w/o DE)   &0.99 & 0.97 & 1.00 & 0.84 & 1.00 & 1.00 & 1.00 & 0.95 & 0.95 & 0.95 & 1.00 & 1.00 & 1.00 & 0.93 & 0.99 & 0.71 & 0.99 & 0.88 & 1.00 & 0.86 & 0.97 & 0.93 & 0.98 & 0.98 & 0.89 & 0.34 & 0.98 & 0.87 \\
        & & LEMON (w/o AP)   &1.00 & 1.00 & 1.00 & 1.00 & 1.00 & 1.00 & 1.00 & 1.00 & 1.00 & 1.00 & 1.00 & 1.00 & 1.00 & 1.00 & 1.00 & 0.99 & 1.00 & 1.00 & 1.00 & 1.00 & 0.99 & 0.99 & 0.98 & 0.98 & 0.98 & 0.73 & 1.00 & 0.98 \\
        & & LEMON (w/o CFG)  &0.96 & 0.96 & 0.94 & 0.84 & 1.00 & 1.00 & 0.97 & 0.80 & 0.91 & 0.91 & 1.00 & 0.97 & 1.00 & 0.68 & 0.79 & 0.38 & 0.91 & 0.72 & 0.94 & 0.77 & 0.93 & 0.75 & 0.93 & 0.92 & 0.78 & 0.28 & 0.93 & 0.77 \\
        & & LEMON            &1.00 & 1.00 & 1.00 & 1.00 & 1.00 & 1.00 & 1.00 & 1.00 & 1.00 & 1.00 & 1.00 & 1.00 & 1.00 & 1.00 & 1.00 & 0.99 & 1.00 & 1.00 & 1.00 & 1.00 & 0.99 & 0.96 & 0.99 & 0.99 & 0.99 & 0.67 & 1.00 & 0.97 \\
        \cmidrule(l){2-31}
        & \multirow{8}{*}{Non-match}
          & LIME             &0.80 & 0.01 & 1.00 & 0.06 & 1.00 & 0.01 & 1.00 & 0.01 & 1.00 & 0.01 & 1.00 & 0.08 & 0.47 & 0.02 & 1.00 & 0.05 & 0.84 & 0.05 & 0.56 & 0.10 & 0.42 & 0.05 & 0.70 & 0.07 & 0.35 & 0.06 & 0.78 & 0.04 \\
        & & SHAP             &0.50 & 0.00 & 1.00 & 0.01 & 1.00 & 0.00 & 1.00 & 0.01 & 0.00 & 0.00 & 1.00 & 0.03 & 1.00 & 0.01 & 1.00 & 0.04 & 0.94 & 0.03 & 0.38 & 0.06 & 0.28 & 0.07 & 0.94 & 0.03 & 0.13 & 0.13 & 0.71 & 0.03 \\
        & & SHAP (w/ CFG)    &1.00 & 0.01 & 1.00 & 0.12 & 1.00 & 0.00 & 1.00 & 0.01 & 0.00 & 0.00 & 1.00 & 0.03 & 1.00 & 0.01 & 1.00 & 0.05 & 1.00 & 0.05 & 0.67 & 0.14 & 0.79 & 0.12 & 1.00 & 0.04 & 1.00 & 1.00 & 0.88 & 0.12 \\
        & & Landmark         &0.16 & 0.13 & 0.85 & 0.83 & 0.51 & 0.08 & 0.45 & 0.13 & 0.36 & 0.17 & 0.22 & 0.19 & 0.96 & 0.90 & 0.04 & 0.04 & 0.44 & 0.41 & 0.43 & 0.36 & 0.04 & 0.03 & 0.81 & 0.77 & 0.09 & 0.09 & 0.41 & 0.32 \\
        & & LEMON (w/o DE)   &0.83 & 0.46 & 0.86 & 0.71 & 0.80 & 0.03 & 0.65 & 0.26 & 0.85 & 0.53 & 0.75 & 0.55 & 0.82 & 0.82 & 0.86 & 0.51 & 0.82 & 0.60 & 0.88 & 0.86 & 0.84 & 0.80 & 0.91 & 0.86 & 0.96 & 0.96 & 0.83 & 0.61 \\
        & & LEMON (w/o AP)   &1.00 & 0.02 & 1.00 & 0.26 & 1.00 & 0.01 & 1.00 & 0.02 & 1.00 & 0.01 & 0.86 & 0.08 & 0.59 & 0.03 & 0.98 & 0.10 & 0.98 & 0.12 & 0.74 & 0.16 & 0.78 & 0.24 & 0.86 & 0.10 & 0.97 & 0.52 & 0.90 & 0.13 \\
        & & LEMON (w/o CFG)  &0.74 & 0.27 & 0.79 & 0.32 & 0.80 & 0.04 & 0.60 & 0.07 & 0.50 & 0.02 & 0.79 & 0.14 & 0.74 & 0.71 & 0.70 & 0.14 & 0.76 & 0.36 & 0.75 & 0.63 & 0.71 & 0.57 & 0.82 & 0.74 & 0.13 & 0.12 & 0.68 & 0.32 \\
        & & LEMON            &0.74 & 0.68 & 0.50 & 0.50 & 0.70 & 0.07 & 0.64 & 0.47 & 0.98 & 0.98 & 0.77 & 0.77 & 0.76 & 0.76 & 0.79 & 0.72 & 0.78 & 0.77 & 0.87 & 0.87 & 0.87 & 0.87 & 0.87 & 0.87 & 0.96 & 0.96 & 0.79 & 0.71 \\
        \midrule
        \multirow{20}{*}{BERT-Mini} & \multirow{10}{*}{Match}
          & LIME             &0.97 & 0.93 & 1.00 & 0.48 & 0.97 & 0.97 & 0.96 & 0.76 & 0.95 & 0.91 & 1.00 & 0.48 & 0.98 & 0.70 & 0.97 & 0.97 & 0.94 & 0.55 & 1.00 & 0.45 & 0.98 & 0.68 & 0.97 & 0.65 & 0.96 & 0.10 & 0.97 & 0.66 \\
        & & SHAP             &0.90 & 0.89 & 0.89 & 0.74 & 0.79 & 0.79 & 0.65 & 0.65 & 0.91 & 0.91 & 0.72 & 0.67 & 0.78 & 0.65 & 0.79 & 0.79 & 0.62 & 0.62 & 0.66 & 0.61 & 0.83 & 0.68 & 0.83 & 0.71 & 0.91 & 0.14 & 0.79 & 0.68 \\
        & & SHAP (w/ CFG)    &0.95 & 0.95 & 0.96 & 0.96 & 0.92 & 0.92 & 0.98 & 0.98 & 0.86 & 0.86 & 0.89 & 0.89 & 0.97 & 0.97 & 0.98 & 0.98 & 0.99 & 0.99 & 1.00 & 1.00 & 0.99 & 0.99 & 1.00 & 1.00 & 1.00 & 0.25 & 0.96 & 0.90 \\
        & & IG               &0.90 & 0.90 & 0.43 & 0.43 & 0.73 & 0.70 & 0.83 & 0.83 & 0.50 & 0.50 & 0.70 & 0.70 & 0.68 & 0.66 & 0.70 & 0.68 & 0.89 & 0.89 & 0.81 & 0.81 & 0.69 & 0.67 & 0.79 & 0.79 & 0.33 & 0.33 & 0.69 & 0.68 \\
        & & IG (w/ CFG)      &0.88 & 0.88 & 0.52 & 0.52 & 0.68 & 0.65 & 0.94 & 0.94 & 0.68 & 0.68 & 0.70 & 0.70 & 0.78 & 0.75 & 0.87 & 0.85 & 0.95 & 0.95 & 0.94 & 0.94 & 0.88 & 0.86 & 0.93 & 0.93 & 0.28 & 0.28 & 0.77 & 0.76 \\
        & & Landmark         &0.98 & 0.98 & 1.00 & 0.87 & 1.00 & 1.00 & 0.94 & 0.94 & 0.86 & 0.86 & 0.96 & 0.93 & 0.89 & 0.80 & 0.99 & 0.99 & 0.83 & 0.83 & 0.90 & 0.90 & 0.91 & 0.85 & 0.92 & 0.76 & 0.09 & 0.08 & 0.87 & 0.83 \\
        & & LEMON (w/o DE)   &1.00 & 0.97 & 1.00 & 0.52 & 1.00 & 1.00 & 1.00 & 0.93 & 1.00 & 0.95 & 1.00 & 0.78 & 0.99 & 0.72 & 1.00 & 1.00 & 1.00 & 0.89 & 1.00 & 0.81 & 0.99 & 0.73 & 1.00 & 0.76 & 1.00 & 0.14 & 1.00 & 0.78 \\
        & & LEMON (w/o AP)   &1.00 & 1.00 & 1.00 & 1.00 & 1.00 & 1.00 & 1.00 & 1.00 & 1.00 & 1.00 & 1.00 & 0.89 & 0.99 & 0.99 & 1.00 & 1.00 & 1.00 & 1.00 & 1.00 & 1.00 & 1.00 & 1.00 & 1.00 & 0.95 & 1.00 & 0.25 & 1.00 & 0.93 \\
        & & LEMON (w/o CFG)  &0.97 & 0.92 & 1.00 & 0.48 & 0.98 & 0.98 & 0.99 & 0.76 & 1.00 & 0.95 & 1.00 & 0.41 & 0.99 & 0.68 & 0.99 & 0.98 & 0.97 & 0.55 & 1.00 & 0.42 & 0.99 & 0.68 & 0.99 & 0.65 & 1.00 & 0.08 & 0.99 & 0.66 \\
        & & LEMON            &1.00 & 1.00 & 1.00 & 1.00 & 1.00 & 1.00 & 1.00 & 1.00 & 1.00 & 1.00 & 1.00 & 0.89 & 0.99 & 0.97 & 1.00 & 1.00 & 1.00 & 0.99 & 1.00 & 1.00 & 1.00 & 1.00 & 1.00 & 0.95 & 1.00 & 0.22 & 1.00 & 0.92 \\
        \cmidrule(l){2-31}
        & \multirow{10}{*}{Non-match}
          & LIME             &0.97 & 0.07 & 1.00 & 0.03 & 1.00 & 0.00 & 1.00 & 0.02 & 1.00 & 0.02 & 0.86 & 0.07 & 0.91 & 0.04 & 1.00 & 0.01 & 1.00 & 0.02 & 1.00 & 0.13 & 0.90 & 0.04 & 0.81 & 0.03 & 0.64 & 0.01 & 0.93 & 0.04 \\
        & & SHAP             &0.90 & 0.07 & 0.86 & 0.09 & 1.00 & 0.01 & 1.00 & 0.02 & 1.00 & 0.01 & 1.00 & 0.17 & 0.78 & 0.04 & 1.00 & 0.01 & 1.00 & 0.02 & 0.84 & 0.21 & 0.70 & 0.08 & 0.59 & 0.08 & 0.76 & 0.10 & 0.88 & 0.07 \\
        & & SHAP (w/ CFG)    &0.95 & 0.08 & 1.00 & 0.15 & 1.00 & 0.01 & 1.00 & 0.02 & 1.00 & 0.01 & 1.00 & 0.21 & 1.00 & 0.06 & 1.00 & 0.01 & 1.00 & 0.03 & 1.00 & 0.32 & 0.96 & 0.09 & 0.92 & 0.23 & 1.00 & 0.15 & 0.99 & 0.10 \\
        & & IG               &0.40 & 0.04 & 0.00 & 0.00 & 0.33 & 0.00 & 0.32 & 0.02 & 0.33 & 0.01 & 0.00 & 0.00 & 0.13 & 0.01 & 0.18 & 0.01 & 0.10 & 0.01 & 0.00 & 0.00 & 0.25 & 0.02 & 0.42 & 0.04 & 0.07 & 0.01 & 0.19 & 0.01 \\
        & & IG (w/ CFG)      &0.67 & 0.04 & 1.00 & 0.01 & 0.33 & 0.00 & 0.48 & 0.02 & 0.33 & 0.01 & 0.00 & 0.00 & 0.44 & 0.02 & 0.29 & 0.01 & 0.21 & 0.01 & 0.17 & 0.01 & 0.70 & 0.03 & 0.65 & 0.04 & 0.62 & 0.02 & 0.45 & 0.02 \\
        & & Landmark         &0.42 & 0.39 & 0.71 & 0.69 & 0.06 & 0.05 & 0.18 & 0.17 & 0.50 & 0.50 & 0.63 & 0.63 & 0.65 & 0.64 & 0.07 & 0.06 & 0.36 & 0.35 & 0.51 & 0.49 & 0.74 & 0.74 & 0.68 & 0.66 & 0.01 & 0.01 & 0.42 & 0.41 \\
        & & LEMON (w/o DE)   &0.99 & 0.61 & 0.95 & 0.91 & 0.87 & 0.41 & 0.93 & 0.54 & 0.92 & 0.81 & 0.94 & 0.89 & 0.98 & 0.88 & 0.96 & 0.61 & 0.95 & 0.67 & 0.91 & 0.86 & 0.99 & 0.96 & 0.95 & 0.91 & 0.97 & 0.97 & 0.95 & 0.77 \\
        & & LEMON (w/o AP)   &1.00 & 0.10 & 1.00 & 0.04 & 1.00 & 0.02 & 1.00 & 0.04 & 1.00 & 0.02 & 1.00 & 0.50 & 1.00 & 0.07 & 1.00 & 0.02 & 1.00 & 0.05 & 1.00 & 0.38 & 1.00 & 0.09 & 1.00 & 0.13 & 1.00 & 0.15 & 1.00 & 0.12 \\
        & & LEMON (w/o CFG)  &1.00 & 0.33 & 1.00 & 0.85 & 1.00 & 0.01 & 0.96 & 0.11 & 0.93 & 0.79 & 1.00 & 0.89 & 0.97 & 0.82 & 1.00 & 0.02 & 0.98 & 0.10 & 0.88 & 0.67 & 0.98 & 0.92 & 0.96 & 0.84 & 0.96 & 0.96 & 0.97 & 0.56 \\
        & & LEMON            &0.82 & 0.81 & 0.94 & 0.94 & 0.80 & 0.55 & 0.69 & 0.67 & 0.86 & 0.86 & 0.98 & 0.96 & 0.90 & 0.90 & 0.52 & 0.49 & 0.79 & 0.79 & 0.87 & 0.87 & 0.95 & 0.95 & 0.98 & 0.98 & 0.97 & 0.97 & 0.85 & 0.83 \\
        \bottomrule
    \end{tabular}
    \label{tab:simulated-cff1-precision-recall}
\end{table*}

\subsection{Magnitude of Changes in User Study}
Table~\ref{tab:user-study-change-magnitude} reports the average edit distance for the record pair,
before and after being altered by the users in the user study (see Section~\ref{sec:user-study}),
after seeing an explanation from LIME or LEMON for all datasets.
We observe that users tend to make bigger changes with LEMON,
perhaps indicating that the users have a tendency to underestimate the changes necessary to sway the matcher when the explanations are less helpful and they need to rely more on their own intuition.
Matches in the Company dataset are a good example.
They require a surprising amount of perturbation to convince the matchers something is not a match because the record pairs contain so many redundant highly discriminative features.

\begin{table}
    \centering
    \caption{
        Average Edit Distance of the Changes Made by the Participants in the User Study.
    }
    \scriptsize
    \begin{tabular}{lcccc}
    \toprule
        \multirow{3}[3]{*}{\textbf{Dataset}} & \multicolumn{4}{c}{\textbf{Method}} \\
        \cmidrule{2-5}
        & \multicolumn{2}{c}{LIME} & \multicolumn{2}{c}{LEMON} \\
        \cmidrule(lr){2-3}
        \cmidrule(lr){4-5}
        & Match & Non-match & Match & Non-match \\
    \midrule
        \textbf{Structured} & & & & \\
        Amazon-Google      &       14 &        15 &         15 &         22 \\
        Beer               &       14 &        20 &         24 &         23 \\
        DBLP-ACM           &       24 &        35 &         43 &         64 \\
        DBLP-GoogleScholar &       20 &        21 &         33 &         39 \\
        Fodors-Zagats      &       14 &        17 &         17 &         16 \\
        iTunes-Amazon      &       17 &        28 &         23 &         35 \\
        Walmart-Amazon     &       12 &        19 &         19 &         17 \\
    \midrule
        \textbf{Dirty} & & & & \\
        DBLP-ACM           &       25 &        25 &         49 &         82 \\
        DBLP-GoogleScholar &       27 &        23 &         47 &         53 \\
        iTunes-Amazon      &       33 &        36 &         28 &         60 \\
        Walmart-Amazon     &       17 &        20 &         26 &         23 \\
    \midrule
        \textbf{Textual} & & & & \\
        Abt-Buy &       14 &        35 &         36 &         36 \\
        Company &      107 &       102 &        556 &        148 \\
    \bottomrule
    \end{tabular}
    \label{tab:user-study-change-magnitude}
\end{table}

\subsection{Neighborhood Sample Size}
\begin{figure*}
    \centering
    \includegraphics{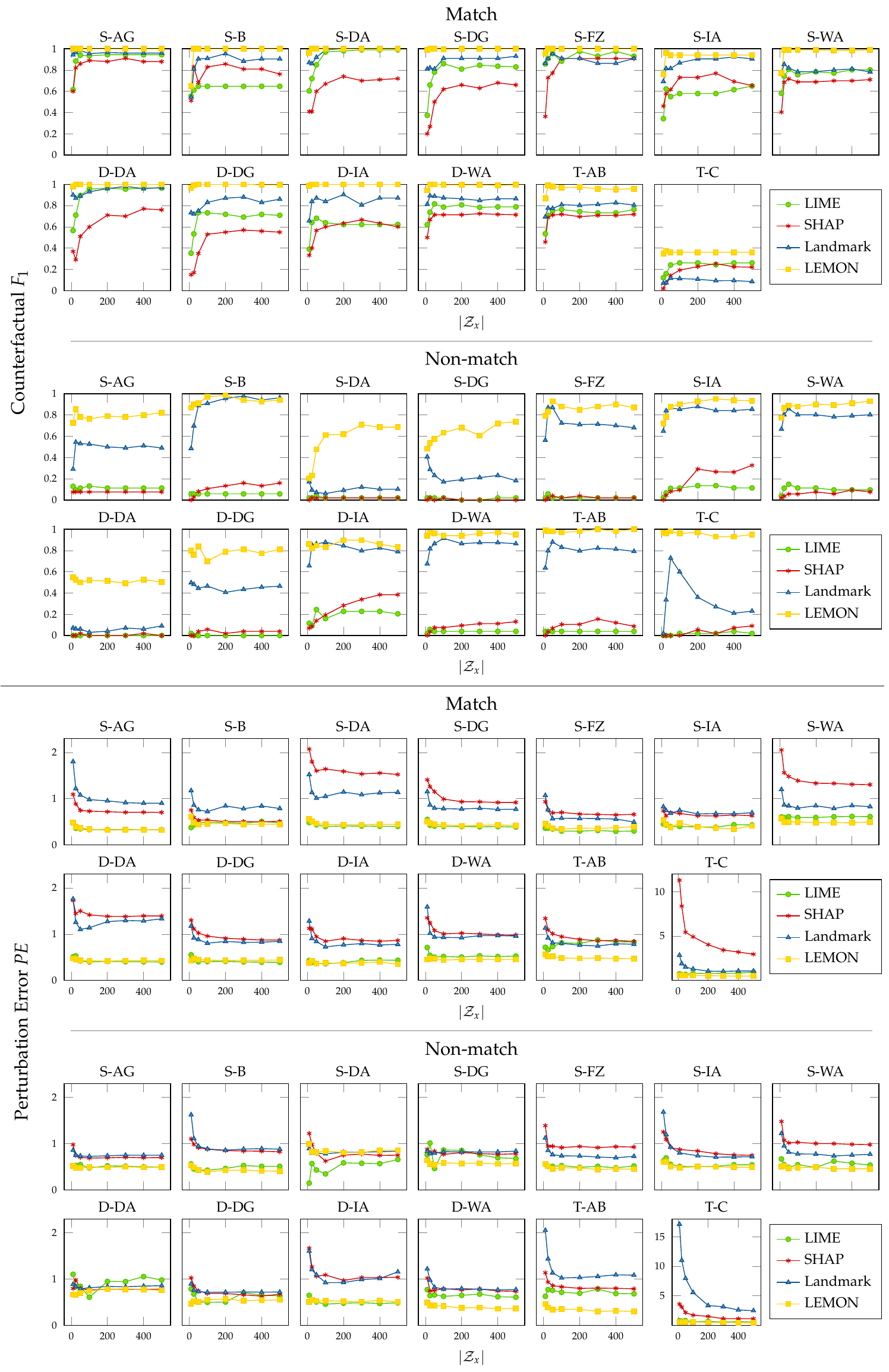}
    \caption{
        Counterfactual $F_1$ and perturbation error ($\mathit{PE}$) of explainability methods based on neighborhood sampling for BERT-Mini on all datasets when varying the neighborhood sampling size $|\mathcal{Z}_x|$.
    }
    \label{fig:vary-num-samples-all}
\end{figure*}

\begin{figure*}
    \centering
    \includegraphics{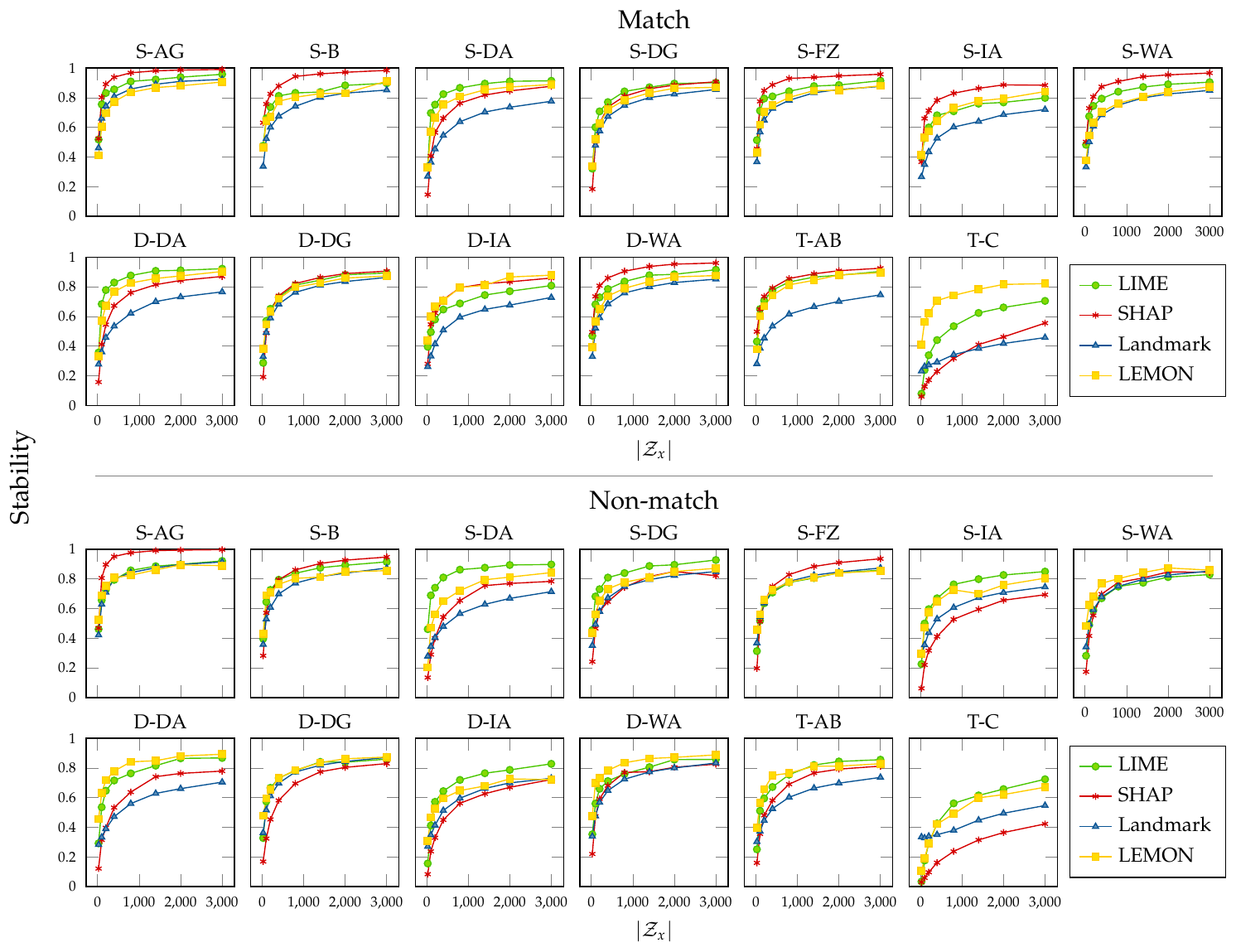}
    \caption{
        Stability of explainability methods based on neighborhood sampling for BERT-Mini on all datasets when varying the neighborhood sampling size $|\mathcal{Z}_x|$.
    }
    \label{fig:stability-vary-num-samples-all}
\end{figure*}

Due to the space constraints, Figure~\ref{fig:vary-num-samples} and \ref{fig:stability-vary-num-samples} from Section~\ref{sec:neighborhood-sample-size} only report results from the Abt-Buy dataset.
Figure~\ref{fig:vary-num-samples-all} and \ref{fig:stability-vary-num-samples-all} show the results for all datasets.

The key takeaway from Section~\ref{sec:neighborhood-sample-size} about neighborhood sampling size $\mathcal{Z}_x$ and performance is true for all datasets:
it takes a relatively low number of samples to reach stationary levels of performance,
and the $F_1$ score and perturbation error do not change much with more samples after that.
We can observe that, unsurprisingly, datasets with larger records tend to need more samples to reach this state.

Overall,
the performance increases with more samples up to a certain point and is significantly hampered by a very low number of samples.
This is not only because the low number of samples leads to erroneous modeling of the effect of perturbations,
but also because there might not have been any interesting perturbations sampled.
However,
we note that in some instances the counterfactual $F_1$ score is higher for a lower number of samples.
For example Landmark on the Company dataset.
Upon inspection,
we see this is because the low number of samples makes the surrogate model overfit and make overly confident claims.
This turns out to be correct more often in a strictly counterfactual sense and pay off in terms of counterfactual $F_1$ score compared to a more faithful approach that fails to provide a counterfactually interpretable explanation.
Unfortunately,
this comes at the cost of unacceptably large perturbation errors and low faithfulness and does therefore not represent a viable option in practice.

In regards to stability,
we see from Figure~\ref{fig:stability-vary-num-samples-all} that the behavior is similar on all datasets.
The main difference is that datasets with bigger records tend to need more samples to reach similar levels of stability.

\subsection{Explanation Complexity}
\begin{figure*}
    \centering
    \includegraphics{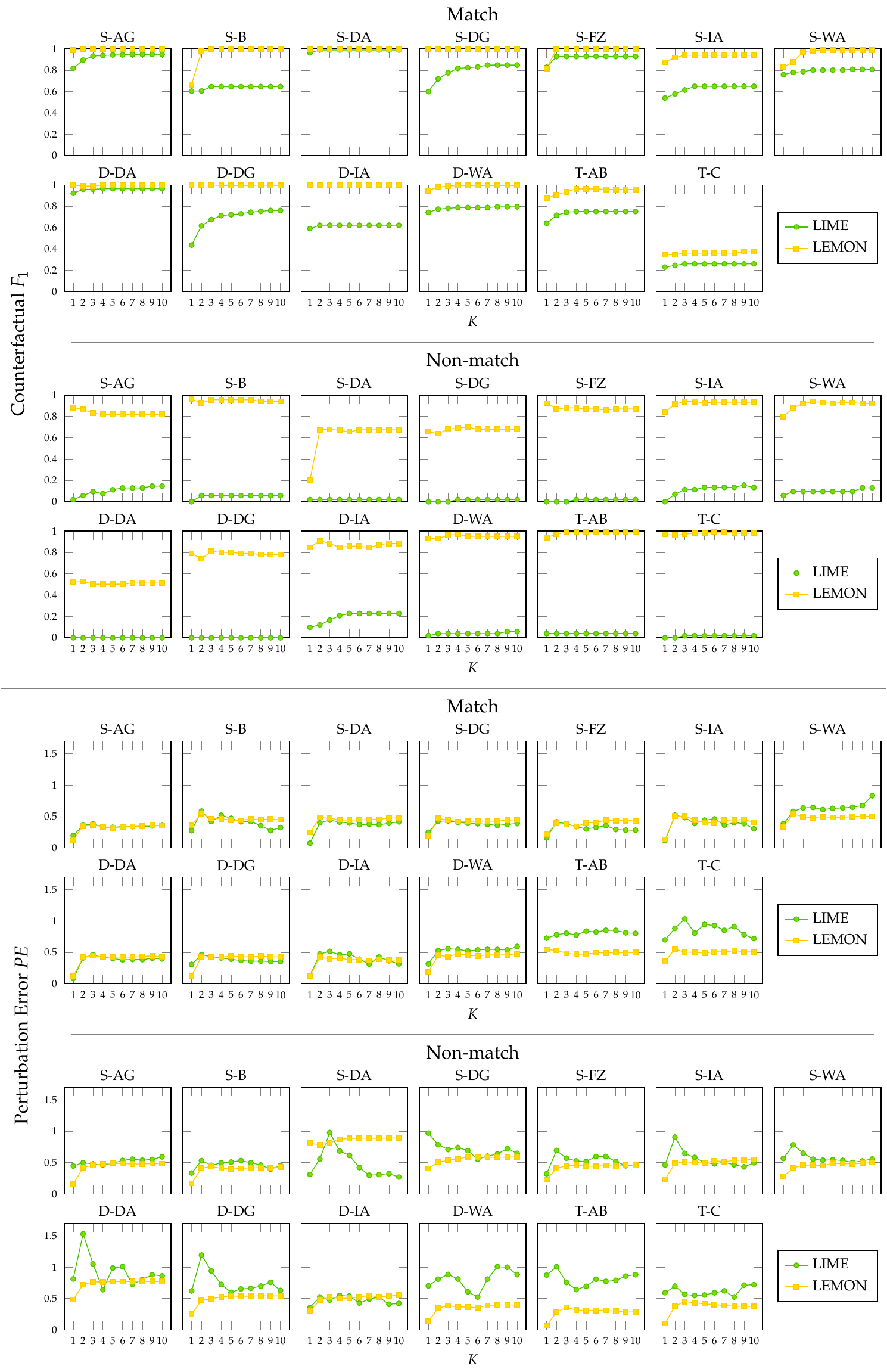}
    \caption{
        Counterfactual $F_1$ and perturbation error ($\mathit{PE}$) of explainability methods based on neighborhood sampling for BERT-Mini on all datasets when varying $K$.
    }
    \label{fig:vary-k-all}
\end{figure*}

Figure~\ref{fig:vary-k} from Section~\ref{sec:explanation-complexity} shows the effect of varying $K$ for the Abt-Buy dataset.
For completeness, Figure~\ref{fig:vary-k-all} shows the effect of varying $K$ for all datasets.
Experiments were performed as explained in Section~\ref{sec:explanation-complexity}.

Results for all datasets verify the claim that,
for all but the lowest of $K$s,
the counterfactual interpretation and explanation faithfulness is not meaningfully affected by the choice of $K$.
This is convenient because it lets us prioritize choosing a $K$ that is suitable for user consumption.

\subsection{Runtime}
\begin{figure*}
    \centering
    \includegraphics{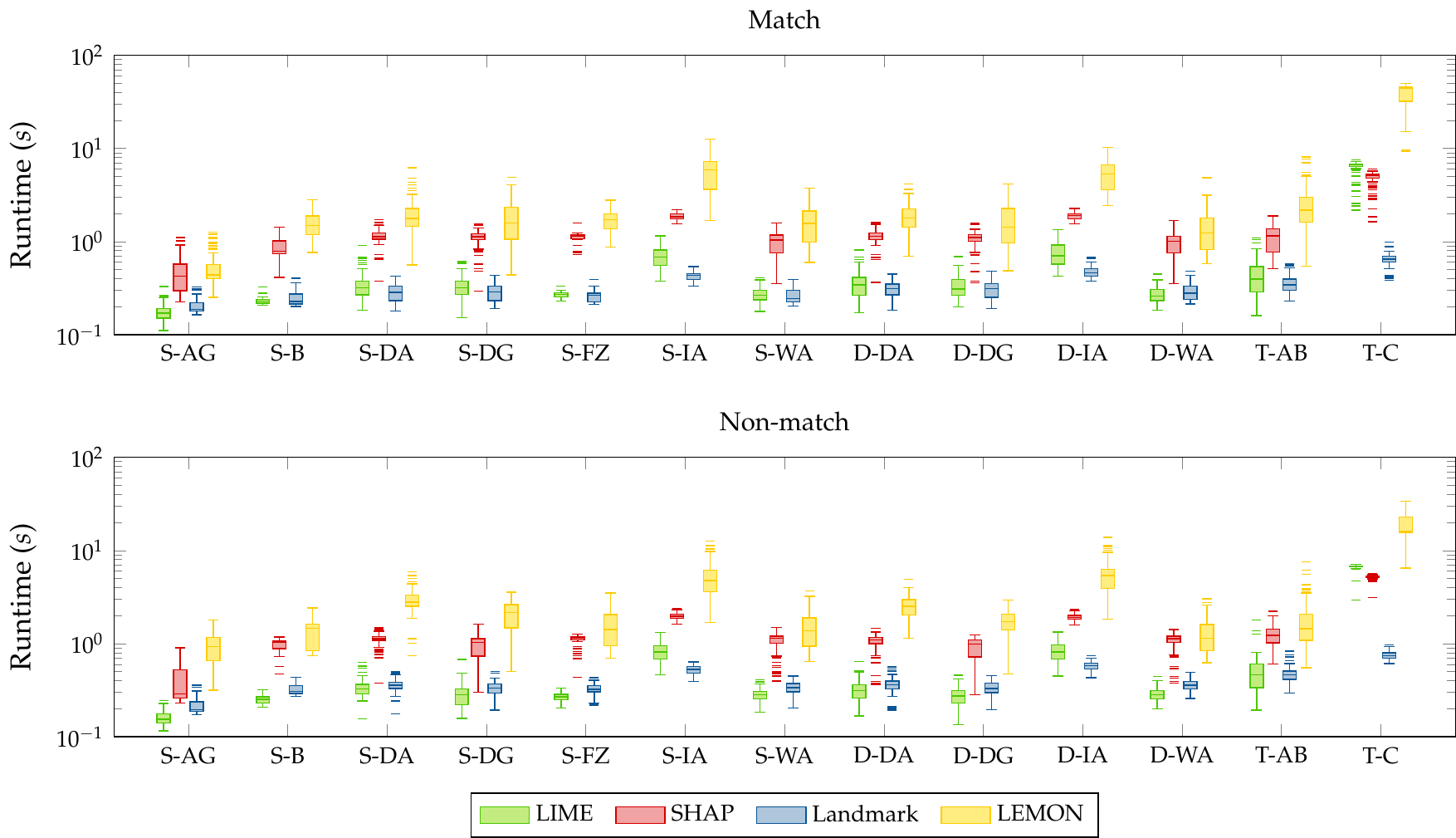}
    \caption{
        Runtime of explainability methods based on neighborhood sampling for BERT-Mini on all datasets.
    }
    \label{fig:runtime-all}
\end{figure*}

\begin{figure*}
    \centering
    \includegraphics{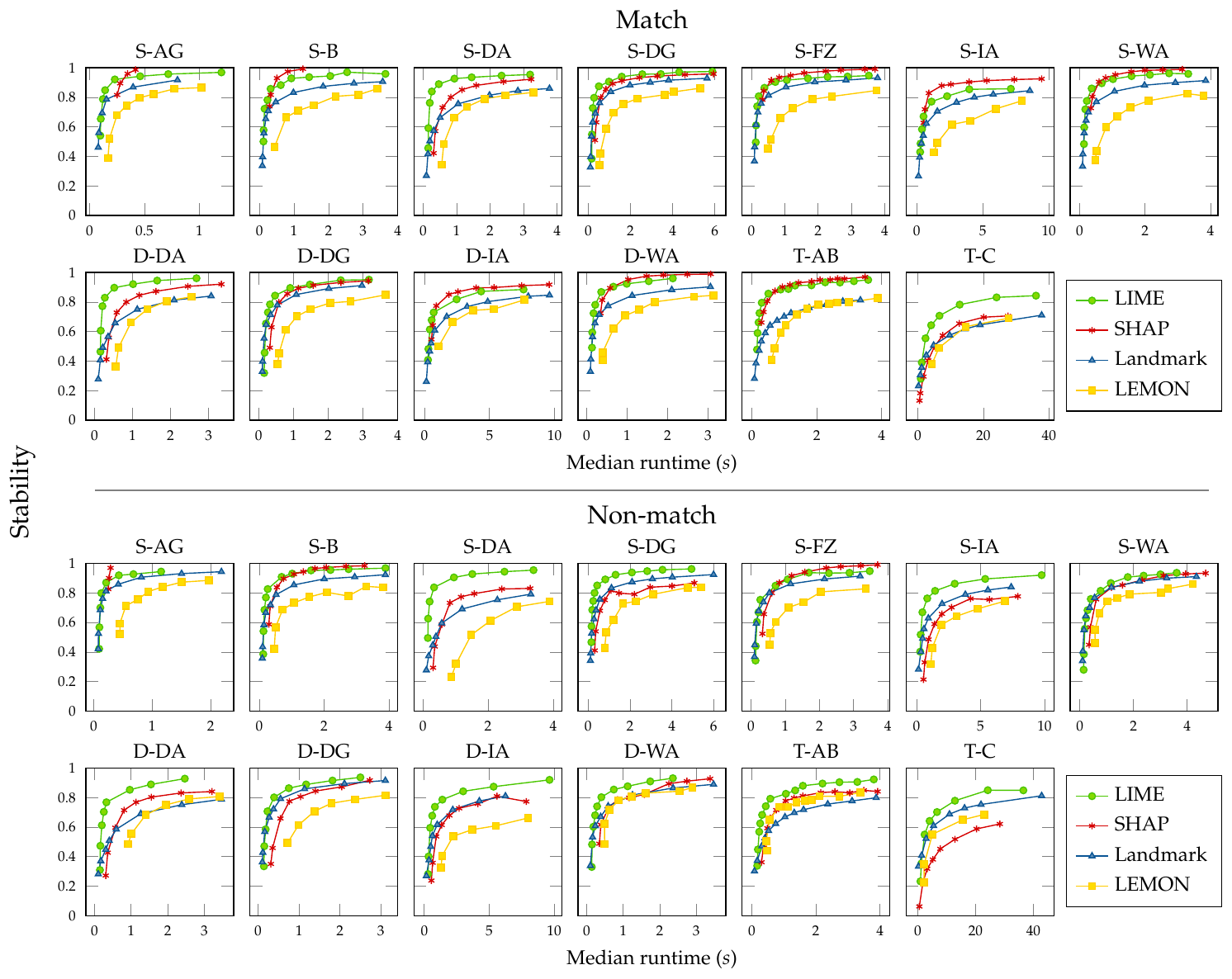}
    \caption{
        Stability of explainability methods based on neighborhood sampling for BERT-Mini on all datasets when varying the median runtime (by changing $|\mathcal{Z}_x|$).
    }
    \label{fig:runtime-stability-all}
\end{figure*}

Figure~\ref{fig:runtime} from Section~\ref{sec:runtime} shows the runtime for three selected datasets.
We report the equivalent results for all datasets in Figure~\ref{fig:runtime-all}.
Furthermore, Figure~\ref{fig:runtime-stability-all} extends the results on stability-runtime trade-off in Figure~\ref{fig:runtime-stability} from Section~\ref{sec:runtime} to all datasets.

In general,
we observe that LEMON has higher runtime than the baselines across all datasets.
The runtime is first and foremost determined by the neighborhood sampling size $\mathcal{Z}_x$.
However,
as discussed in Section~\ref{sec:stability} and \ref{sec:neighborhood-sample-size},
even small sample sizes yield satisfactory counterfactual interpretation and explanation faithfulness,
and deciding $\mathcal{Z}_x$ in practice is mostly a matter of stability.
Therefore,
if low runtime is important,
one has the option to trade off some stability to decrease the runtime.
Figure~\ref{fig:runtime-stability-all} then tells a different story than Figure~\ref{fig:runtime-all} because it shows that the trade-off between runtime and stability is less than the relative difference in runtime as seen in Figure~\ref{fig:runtime-all} for most datasets.
In other words,
one can decrease the neighborhood sampling size of LEMON to get a more similar runtime as for example Landmark while still being equally stable and retaining the high level of counterfactual interpretation and explanation faithfulness.
To what degree this trade-off is beneficial depends on the dataset.

\end{document}